\documentclass[manuscript,screen,acmlarge,11pt]{acmart}

\AtBeginDocument{%
  \providecommand\BibTeX{{%
    Bib\TeX}}}

\setcopyright{acmlicensed}
\copyrightyear{2018}
\acmYear{2018}
\acmDOI{XXXXXXX.XXXXXXX}

\acmConference[Conference acronym 'XX]{Make sure to enter the correct
  conference title from your rights confirmation emai}{June 03--05,
  2018}{Woodstock, NY}
\acmISBN{978-1-4503-XXXX-X/18/06}

\usepackage{algorithmic}
\usepackage{graphicx}
\usepackage{textcomp}
\usepackage{xcolor}
\usepackage{url}
\usepackage{booktabs} 
\usepackage{multirow}
\usepackage{tabularx}
\usepackage{subcaption}
\usepackage{makecell}
\usepackage{framed}
\usepackage{enumitem}
\usepackage{xcolor}
\usepackage[most]{tcolorbox}
\usepackage{colortbl}
\usepackage{diagbox}
\usepackage{hyperref}
\usepackage[switch]{lineno}
\newcolumntype{C}{>{\centering\arraybackslash}X}
\newcommand{\MYhref}[3][blue]{\href{#2}{\color{#1}{#3}}}%

\def\BibTeX{{\rm B\kern-.05em{\sc i\kern-.025em b}\kern-.08em
    T\kern-.1667em\lower.7ex\hbox{E}\kern-.125emX}}

\definecolor{dark-red}{RGB}{255,0,0}
\definecolor{dark-green}{RGB}{0,200,0}
\definecolor{dark-blue}{RGB}{0,0,139}

\begin{document}

\title{CodeUltraFeedback: An LLM-as-a-Judge Dataset for Aligning Large Language Models to Coding Preferences}

\author{Martin Weyssow}
\email{martin.weyssow@umontreal.ca}
\affiliation{%
  \institution{DIRO, Universit\'e de Montr\'eal}
  \city{Montreal}
  \country{Canada}
}

\author{Aton Kamanda}
\email{atonkamanda@hotmail.com}
\affiliation{%
  \institution{DIRO, Universit\'e de Montr\'eal}
  \city{Montreal}
  \country{Canada}
}

\author{Xin Zhou}
\email{xinzhou.2020@phdcs.smu.edu.sg}
\affiliation{%
  \institution{Singapore Management University}
  \city{Singapore}
  \country{Singapore}
}

\author{Houari Sahraoui}
\email{sahraouh@iro.umontreal.ca}
\affiliation{%
  \institution{DIRO, Universit\'e de Montr\'eal}
  \city{Montreal}
  \country{Canada}
}

\renewcommand{\shortauthors}{Weyssow et al.}

\begin{abstract}
Evaluating the alignment of large language models (LLMs) with user-defined coding preferences is a challenging endeavour that requires a deep assessment of LLMs' outputs.
Existing methods and benchmarks rely primarily on automated metrics and static analysis tools, which often fail to capture the nuances of user instructions and LLM outputs.
To address this gap, we propose using the LLM-as-a-Judge methodology to evaluate the alignment of LLMs with coding preferences. 
Based on this approach, we present CodeUltraFeedback, a comprehensive dataset designed to facilitate the evaluation and improvement of LLM alignment.
CodeUltraFeedback consists of 10,000 coding instructions, each annotated with four responses generated from a diverse pool of 14 LLMs. 
These responses are ranked based on five distinct coding preferences using GPT-3.5 as a judge, providing both numerical scores and detailed textual feedback.
Our analysis of CodeUltraFeedback reveals that responses from GPT-3.5 and GPT-4 are generally preferred over those from open-weight LLMs, highlighting significant differences in alignment between closed and open-weight models.
In turn, we explore the usage of CodeUltraFeedback as feedback data to fine-tune and align CodeLlama-7B-Instruct using supervised fine-tuning (SFT) and reinforcement learning from AI feedback (RLAIF) with direct preference optimization (DPO). 
The resulting aligned CodeLlama-7B-Instruct model outperforms larger LLMs in terms of alignment with coding preferences and shows improved functional correctness on the HumanEval+ benchmark compared to the original instruct model.
Therefore, our contributions bridge the gap in preference tuning of LLMs for code and set the stage for further advancements in model alignment and RLAIF in automated software engineering.
\end{abstract}



\maketitle

\section{Introduction}
The advent of recent large language models (LLMs) has ushered in a new era of LLMs with high coding capabilities~\cite{chen2021evaluating, roziere2023code, guo2024deepseek, jiang2023mistral, touvron2023llama, muennighoff2023octopack}, showcasing remarkable performances across a wide range of downstream tasks including code generation~\cite{ liu2023your, athiwaratkun2022multi, weyssow2023exploring, zheng2023codegeex}, code translation~\cite{jiao2023evaluation, pan2023understanding}, bug fixing~\cite{silva2023repairllama, muennighoff2023octopack, ye2021comprehensive}, and more.
However, as LLMs continue to advance, a critical question emerges: \textit{How well do these capabilities align with developers' expectations, particularly regarding non-functional requirements such as code readability, efficiency, and adherence to best practices?}

Current methodologies for fine-tuning and evaluating LLMs primarily focus on core capabilities, e.g., translating, summarizing, or reviewing code~\cite{lu2021codexglue, puri2021codenet, zhu2022xlcost, niu2023crosscodebench, zhou2023generation, sghaier2024improving, sghaier2023unity, sghaier2023multi}, and functional correctness in diverse code generation scenarios~\cite{hendrycksapps2021, austin2021program, chen2021evaluating, muennighoff2023octopack, liu2023your, khan2023xcodeeval}.
While some research has started to address non-functional properties of LLM-generated code, such as code quality~\cite{siddiq2023lightweight}, runtime efficiency~\cite{huang2024effibench}, and other aspects~\cite{yeticstiren2023evaluating, singhal2024nofuneval}, these efforts often fall short in several key areas. 
First, existing benchmarks typically refine established metrics without considering more open-ended coding problems that involve complex user instructions. 
Second, the assessment of LLM outputs often relies on automated metrics and external tools based on inflexible standards and patterns. 
This approach fails to capture the nuanced complexity of users' instructions and the subtleties in LLMs' responses.
Third, there is a notable absence of large-scale datasets specifically designed for tuning and aligning LLMs to non-functional requirements.

The landscape of existing datasets and benchmarks in automated software engineering reveals a critical gap: the lack of methods for both tuning LLMs and accurately measuring their alignment with non-functional requirements.
Furthermore, current evaluation methods fall short of capturing the nuances in natural and programming languages, underscoring the need for a more sophisticated, language-centric approach to assessment.
To address these shortcomings, we must develop comprehensive datasets designed to improve LLM alignment with non-functional requirements and design evaluation methods capable of evaluating LLM alignment with greater precision.
By closing these gaps, we can better bridge the divide between developers' expectations and the output of LLMs, ensuring that the generated code not only functions correctly but also adheres to best practices, maintains readability, and operates efficiently.

In this paper, we introduce two novel contributions: \textit{CodeUltraFeedback} and \textit{CodeUltraFeedback-Bench}.
CodeUltraFeedback is a preference dataset comprising 10,000 complex instructions with 40,000 meticulously curated LLMs responses aligned with five non-functional requirements (or \textit{coding preferences}\footnote{In the rest of this paper, we use the term ``coding preferences'' instead of ``non-functional requirements'' to more precisely capture the process of instructing LLMs with explicit preferences, although both terms bear similar meanings.}): instruction following, code explanation, code complexity and efficiency, code readability, and coding style.
The objective of CodeUltraFeedback is to serve as a dataset for preference tuning of LLMs, leveraging recent advancements in LLM alignment, including UltraFeedback~\cite{cui2023ultrafeedback}, reinforcement learning from AI feedback (RLAIF)~\cite{bai2022constitutional, lee2023rlaif, ouyang2022training} and LLM-as-a-Judge~\cite{zheng2023judging, alpaca_eval}, predicated upon the advanced judging capabilities of LLMs like GPT-3.5 or GPT-4.
We adopt an approach auxiliary to UltraFeedback for building our dataset, starting by tagging each instruction with a coding preference.
Then, we generate responses to the instructions using four LLMs randomly selected from a pool of 14 LLMs to achieve diversity and consider various writing styles.
Finally, we use LLM-as-a-Judge with GPT-3.5 to rate the alignment of the LLM responses with the instruction's coding preference, resulting in annotations comprising a numerical rating and a textual rationale for the rating.
The resulting dataset can then be leveraged to align an LLM to generate high-quality responses given a coding preference by tuning the LLM through RLAIF using techniques like direct preference optimization (DPO)~\cite{rafailov2023direct}.

Subsequently, we construct CodeUltraFeedback-Bench by selecting a subset of 500 instructions from CodeUltraFeedback.
CodeUltraFeedback-Bench serves a different purpose than CodeUltraFeedback and aims to comprehensively assess and compare the alignment of LLMs with coding preferences. 
We design a rigorous single-answer grading scheme using LLM-as-a-Judge that allows the judge LLM to evaluate LLM-generated code consistently and objectively.
This approach offers a more nuanced evaluation strategy than previous methods that rely on automated metrics and external tools by leveraging the advanced reasoning capabilities of LLMs. 
Their ability to discern nuances in language allows for a more refined and context-sensitive evaluation of how well LLM-generated code aligns with coding preferences in a language-centric fashion.

An initial data analysis of CodeUltraFeedback's annotations reveals a lack of alignment of 12 LLMs, including strong LLMs such as WizardCoder-33B~\cite{luo2023wizardcoder} and DeepSeek-Coder-Instruct-33B~\cite{guo2024deepseek}, while GPT-3.5 and GPT-4 demonstrate strong alignment.
In turn, we explore preference tuning of a small model, CodeLlama-7B-Instruct~\cite{roziere2023code} using CodeUltraFeedback with supervised fine-tuning (SFT) and RLAIF with DPO~\cite{rafailov2023direct, tunstall2023zephyr}.
This method selects high and low-rated responses to align CodeLlama-7B-Instruct with coding preferences using DPO, encouraging the model to favor highly rated responses.
We demonstrate that tuning CodeLlama-7B-Instruct with SFT and DPO substantially improves alignment across all coding preferences, surpassing all larger LLMs, including CodeLlama-34B-Instruct and WizardCoder-33B when using GPT-3.5-Turbo as a judge.
Further, we explore LLM alignment on CodeUltraFeedback-Bench using GPT-4-Turbo as a judge and highlight more mitigated effects of SFT and DPO, albeit CodeLlama-7B-Instruct still outperforms CodeLlama-13B-Instruct and CodeLlama-34B-instruct in this setup.
We also show that preference tuning does not hinder CodeLlama-7B-Instruct's ability to generate functionally correct code, substantially improving Pass@$k$ on HumanEval~\cite{chen2021evaluating} and HumanEval+~\cite{liu2023your}.
Finally, we implement QLoRA~\cite{dettmers2023qlora} for efficient fine-tuning, showing that SFT and DPO can effectively operate on a single RTX A5000 GPU (24GB). 
Our work establishes strong evidence of the benefits of SFT and DPO in improving LLM alignment with coding preferences, demonstrating advancements in both preference alignment and functional correctness. This paves the way for future improvements in LLM training and evaluation, promising closer alignment with developers' nuanced expectations.

To summarize, our main contributions are the following:
\begin{itemize}[leftmargin=*]
    \item[--] We release CodeUltraFeedback, a preference dataset of 10,000 complex instructions and 40,000 responses generated using 14 diverse LLMs for aligning LLMs to coding preferences in a code generation scenario.

    \item[--] We contribute CodeUltraFeedback-Bench, a benchmark to evaluate and compare LLM alignment using LLM-as-a-Judge over five coding preferences: instruction following, code explanation, code complexity and efficiency, code readability, and coding style.

    \item[--] We validate the utility of CodeUltraFeedback in facilitating LLM alignment of CodeLlama-7B-Instruct using SFT and DPO.
\end{itemize}

Our models, dataset, benchmark, and prompt templates are available at \MYhref[dark-blue]{https://github.com/martin-wey/CodeUltraFeedback}{https://github.com/martin-wey/CodeUltraFeedback}.

\section{CodeUltraFeedback}

\begin{figure}[!t]
    \centering
    \includegraphics[width=.9\linewidth]{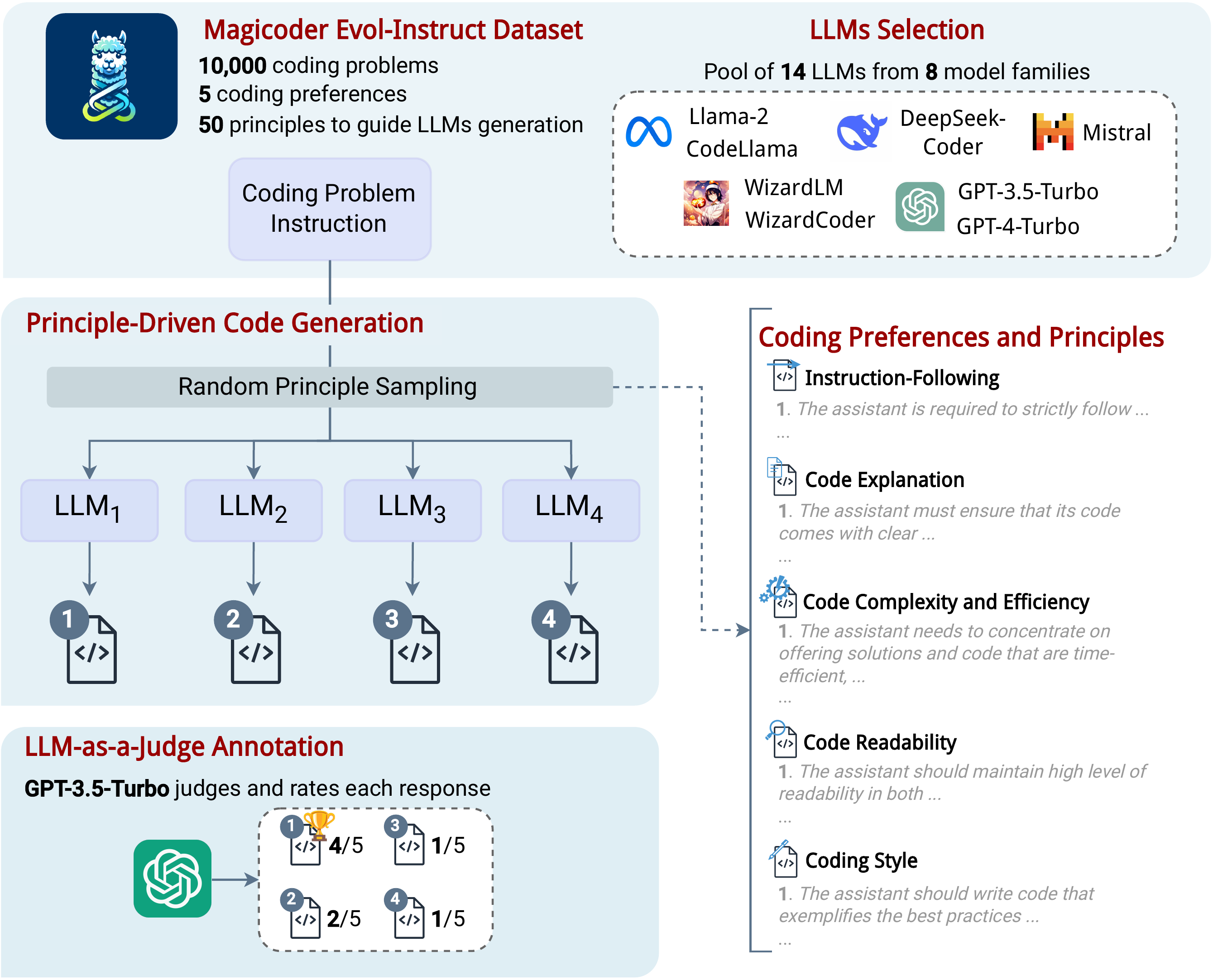}
    \caption{Overview of CodeUltraFeedback dataset construction procedure. (i) \textit{Coding Preferences and Principles}: we define five coding preferences: Instruction-following, Code Explanation, Code Complexity and Efficiency, Code Readability, and Coding Style, and 10 principles for each preference to guide LLMs' generative process. 
    (ii) \textit{Magicoder Evol-Instruct Dataset}: CodeUltraFeedback is based on a 10k-instruction subset of Magicoder Evol-Instruct dataset~\cite{wei2023magicoder, luo2023wizardcoder}, where each instruction is tagged with a coding preference.
    (iii) \textit{LLMs Selection}: for each instruction, four LLMs are randomly sampled from a diverse pool of 14 LLMs. 
    (iv) \textit{Principle-Driven Code Generation}: one principle is randomly selected per LLM to guide and align the code generation process with the assigned coding preference.
    (v) \textit{LLM-as-a-Judge Annotation}: LLM-as-a-Judge with GPT-3.5-Turbo is used to judge LLMs' responses according to coding preferences evaluation criteria (see Table~\ref{tab:readability_judging_template}).}
    \label{fig:approach}
\end{figure}

In this section, we introduce our new dataset, CodeUltraFeedback.
Fig.~\ref{fig:approach} depicts our methodology for building CodeUltraFeedback, which incorporates various components and ideas that we describe in the subsequent subsections.

\subsection{Coding Preferences and Principles}
We start by identifying five coding preferences to guide the creation of our dataset and which are essential to evaluate the broader capabilities of LLMs: (1) \textbf{Instruction-Following} is about the strict adherence of the LLM to the instructions provided by users. 
This preference is foundational for ensuring that LLMs truly follow the user intent and thus provide personalized responses to instructions. 
(2) \textbf{Code Explanation} emphasizes generating code with detailed natural language explanations. 
It underscores the importance of an LLM in providing a solution with explanations that can serve as a bridge between potentially complex code and users while improving trustworthiness.
(3) \textbf{Code Complexity and Efficiency} preference underlines the LLM capability to generate code optimized for performance in terms of speed and resource utilization. 
It is another crucial aspect requiring the LLM to balance the speed and resource usage of the solution carefully. 
(4) \textbf{Code Readability} serves a distinct purpose to the code explanation preference by emphasizing the clarity and understandability of the code itself through its structure, style, and the presence of meaningful documentation and in-line comments. 
(5) \textbf{Coding Style} preference focuses on the importance of writing code that not only meets syntactical correctness but also aligns with the idiomatic practices and stylistic norms of the programming language.

Altogether, these five preferences complement the functional correctness requirement of LLM-generated code. 
Assessing how well LLMs align with these preferences amounts to evaluating how the generated solutions are optimized for human comprehension, maintainability, performance, and collaborative work.

\subsection{Initial Dataset} 
To build CodeUltraFeedback, we rely on Magicoder Evol-Instruct dataset~\cite{wei2023magicoder}, an Evol-Instruct~\cite{xu2023wizardlm, luo2023wizardcoder} version of CodeAlpaca dataset comprising coding problem instructions. 
We select a random subset of Magicoder Evol-Instruct comprising 10,000 samples serving as an initial dataset for CodeUltraFeedback. 
We assign 2,000 samples per coding preference, ensuring a balanced representation of each preference in the dataset.
The reason for subsampling the dataset to 10,000 samples is to lower costs related to OpenAI API needed to generate some of the responses and the annotations. 

\subsection{LLMs Selection}
In order to generate highly diverse solutions to the instructions, we select a pool of 14 LLMs spanning eight model families. 
These include GPT-4-Turbo and GPT-3.5-Turbo as closed-source LLMs. 
We include Llama-2-13B-Chat and Llama-2-70B-Chat~\cite{touvron2023llama}, specialized for chat use cases. 
We choose CodeLlama-7/13/34B-Instruct models~\cite{roziere2023code}, instruction-tuned for code generation. 
We also include WizardLM-15/33B models~\cite{xu2023wizardlm} and Mistral-7B-Instruct~\cite{jiang2023mistral}, which are instruction-tuned LLMs that can follow complex instructions. 
Finally, we include WizardCoder-15/33B~\cite{luo2023wizardcoder} and DeepSeek-Coder-6.7/33B-Instruct~\cite{guo2024deepseek}, recent instruction-tuned LLMs that demonstrate state-of-the-art performance on code generation benchmarks such as HumanEval~\cite{chen2021evaluating} and HumanEval+~\cite{liu2023your}.
From this pool, we randomly select four LLMs for each instruction to generate responses, allowing for the representation of a variety of coding approaches and styles.
By including models that encompass diverse pre-training and instruction-tuning phases, our dataset captures a wide spectrum of response generation behaviors. 
The differences between LLMs extend to pre-training objectives, with some models specializing in general natural language tasks (e.g., Llama-2-13B-Chat) and others explicitly optimized for code-related tasks (e.g., WizardCoder and DeepSeek-Coder).
This diversity ensures that the dataset reflects not only distinct model capabilities but also various stylistic and structural approaches to solving programming problems.
In turn, this strategy prevents subsequent fine-tuning processes that use the CodeUltraFeedback preference dataset to overfit a specific style of response.

\subsection{Principle-Driven Code Generation} 
\label{sec:principle_code_gen}
In this step, we leverage principle-driven generation and generate 10 principles per coding preference using ChatGPT following prior work practices~\cite{sun2023principle, cui2023ultrafeedback, mukherjee2023orca}. 
The rationale for establishing 10 principles per preference is to diversify the dataset.
All principles for each coding preference can be found in Tables~\ref{tab:principles_instruction-following}--\ref{tab:principles_style}.

During code generation, one principle, corresponding to the coding preference of the sample, is randomly chosen per LLM and appended to the instruction in an input prompt to guide the generation.
This strategy helps the LLMs generate responses that potentially align with the requirements of the coding preferences. 
The output of this step is the four LLMs' responses to the principle-augmented instructions. 

\subsection{LLM-as-a-Judge Annotation}
\label{sec:llm_annotation}

\renewcommand{\arraystretch}{1}
\setlength{\arrayrulewidth}{.5pt}
\setlength{\tabcolsep}{3pt}
\begin{table}[!t]
\centering
\footnotesize
\caption{Code readability assessment template.}
\label{tab:readability_judging_template}
\vspace{-1em}
    \begin{tabular*}{\linewidth}{p{.99\linewidth}}
    \toprule
Evaluate the readability of code segments. Assess how comments and documentation contribute to understanding the code's logic, purpose, and operation.

\textbf{Evaluation Criteria}: \\
\:\: \textit{Clarity}: How clear and understandable are the code and its accompanying comments/documentation? \\
\:\: \textit{Conciseness}: Are the comments and documentation succinct yet informative? \\
\:\: \textit{Relevance}: Do the comments and documentation directly contribute to explaining the code's logic, objectives, and functionality? \\
\:\: \textit{Comprehensibility}: Can users of varying technical backgrounds easily grasp the code's purpose and how it works? \\
\textbf{Scoring}: Rate outputs on a scale of 1 to 5: \\
\:\:1. \textit{Poor Readability}: The code is hard to follow, with little to no helpful comments/documentation. \\
\:\:2. \textit{Basic Readability}: The code has minimal comments/documentation, offering limited clarity or insight. \\
\:\:3. \textit{Good Readability}: The code is reasonably clear with comments/documentation that aid understanding, though some areas could be improved. \\
\:\:4. \textit{Very Good Readability}: The code and comments/documentation are clear and concise, making the code's logic and purpose easily understandable. \\
\:\:5. \textit{Excellent Readability}: The code exemplifies outstanding readability, with clear, concise, and comprehensive comments/documentation that make it accessible to all users. \\

    \arrayrulecolor{black}
    \bottomrule
    \end{tabular*}
\end{table}

We apply the LLM-as-a-Judge methodology~\cite{zheng2023judging} to annotate responses using GPT-3.5-Turbo. 
Drawing from practices established in previous work~\cite{cui2023ultrafeedback}, we implement a set of guidelines ensuring GPT-3.5-Turbo delivers both consistent and objective feedback.
Each annotation consists of (1) \textbf{a score on a 1--5 scale} offering a comprehensive assessment of the response's alignment, and (2) \textbf{a rationale for the assigned score}, highlighting the response's strengths and identifying any potential shortcomings. 
Tasking GPT-3.5-Turbo to provide a rationale for the scores encourages a chain-of-thought~\cite{wei2022chain} process during the generation of annotations, elucidating the reasons behind the scores and meaningful feedback.
In addition, this process mitigates the risk of variability and subjectivity that might arise from purely quantitative assessments.
Consequently, this reflective process guarantees that the scoring is not just a surface-level judgment but can be backed by thoughtful consideration of some well-defined coding preferences evaluation criteria.

\renewcommand{\arraystretch}{1}
\setlength{\arrayrulewidth}{.5pt}
\setlength{\tabcolsep}{3pt}
\begin{table}[!t]
\centering
\footnotesize
\caption{Excerpt of ratings and rationales generated by GPT-3.5-Turbo using LLM-as-a-Judge prompting. GPT-3.5-Turbo generates clear rationales for each rating demonstrating its strong judging capabilities.}
\label{tab:ratings_overview}
\vspace{-1em}
    \begin{tabular*}{\linewidth}{p{.99\linewidth}}
    \toprule
    {\textbf{Instruction}: Devising a Python-based algorithm that utilizes the technique of ternary search, could you draw this up while also illustrating the functionality of each critical stage in the searching process?} \\
    \textbf{Preference}: Readability \\
    \midrule
    \arrayrulecolor{black!50}
    \textbf{GPT-4}: \texttt{[Response]} \\
    \textbf{Rating}: 5 \\
    \textbf{Rationale}: {\color{black!60}The code is accompanied by clear and concise comments that comprehensively explain each critical stage of the ternary search algorithm. The comments provide a thorough understanding of the algorithm's logic, purpose, and functionality, making it easily accessible to users of varying technical backgrounds. Additionally, the comments include a detailed explanation of the algorithm's relevance and efficiency compared to binary search, enhancing the overall comprehensibility of the code.} \\
    \midrule
    \textbf{DeepSeek-Coder-6.7B-Instruct}: \texttt{[Response]} \\
    \textbf{Rating}: 4 \\
    \textbf{Rationale}: \texttt{[Rationale]} \\
    \midrule
    \textbf{Mistral-7B-Instruct}: \texttt{[Response]} \\
    \textbf{Rating}: 3 \\
    \textbf{Rationale}: \texttt{[Rationale]} \\
    
    \midrule
    \textbf{CodeLlama-13B-Instruct}: \texttt{[Response]} \\
    \textbf{Rating}: 2 \\
    \textbf{Rationale}: {\color{black!60}The code is clear, but the comments are minimal and lack sufficient detail to fully explain the ternary search algorithm and its critical stages. While the comments briefly describe the function's purpose, they do not provide a comprehensive explanation of the algorithm's logic and functionality. Additionally, the comments do not cover the relevance of ternary search in algorithmic strategies, making it less accessible to users with varying technical backgrounds.} \\
    \arrayrulecolor{black}
    \bottomrule
    \end{tabular*}
\end{table}

Table~\ref{tab:readability_judging_template} illustrates the assessment template and evaluation criteria for code readability evaluation.
It aims to provide precise and comprehensive criteria to the LLM judge for response assessment.
We design analogous assessment templates, each tailored with specific evaluation criteria, for evaluating the other coding preferences (see Tables~\ref{tab:instruction_judging_template}--\ref{tab:style_judging_template}).
The annotation procedure entails constructing a prompt combining the coding preference assessment template and the four LLMs responses.
GPT-3.5-Turbo is then tasked to generate the annotations for all four responses simultaneously, enhancing the consistency and reliability of the evaluation process over individual response assessments.
Table~\ref{tab:ratings_overview} presents an example of the output of the annotation process using GPT-3.5-Turbo, with the LLMs responses and some rationales omitted for brevity.
This example showcases GPT-3.5-Turbo's capability to provide comprehensive justifications for each rating, demonstrating its effective evaluation capabilities.
Furthermore, Table~\ref{tab:ratings_overview} illustrates an example of relatively complex instructions included in CodeUltraFeedback, showcasing the complexity of the task.

\section{CodeUltraFeedback: Dataset Analysis}
\renewcommand{\arraystretch}{1.1}
\setlength{\arrayrulewidth}{.5pt}
\begin{table*}[!t]
\centering
\small
\caption{Average scores across coding preferences for each LLM. The statistical significance of differences in ratings between each LLM and GPT-3.5-Turbo was tested using pairwise Welch's $t$-tests. Significance levels are denoted as follows: * for $p < 0.05$, $\dagger$ for $p < 0.01$, and $\ddagger$ for $p < 0.001$. The results from almost all pairwise $t$-tests indicate highly significant differences ($p < 0.001$).} 
\label{tab:initial_scores}
\vspace{-1em}
\resizebox{.9\textwidth}{!}{%
\rowcolors{2}{gray!15}{white}
    \begin{tabular}{l|ccccc|c}
    \toprule
    Model & {\makecell[c]{Instruction \\ Following}}
    
     & {\makecell[c]{Code \\ Explanation}} & {\makecell[c]{Code Complexity \\ \& Efficiency}} & {\makecell[c]{Code \\ Readability}} & {\makecell[c]{Coding \\ Style}} & Average \\
    \midrule
    GPT-4-Turbo & $\mathbf{3.79}^{\ddagger}$ & $\mathbf{4.04}^{\ddagger}$ & $\mathbf{3.91}^{\dagger}$ & $\mathbf{4.14}^{\ddagger}$ & $\mathbf{4.03}^{\ddagger}$ & \textbf{3.98} \\
    
    GPT-3.5-Turbo & \underline{3.56} & \underline{3.76} & \underline{3.76} & \underline{3.71} & \underline{3.66} & \underline{3.69} \\

    \arrayrulecolor{black!50}
    \midrule
    
    WizardCoder-33B & $3.29^{\ddagger}$  & $3.31^{\ddagger}$ & $3.43^{\ddagger}$  & $3.44^{\ddagger}$ & $3.49^{\dagger}$  & 3.39 \\
    
    DeepSeek-Coder-33B-Instruct & $3.31^{\ddagger}$  & $3.30^{\ddagger}$  & $3.32^{\ddagger}$  & $3.46^{\ddagger}$  & $3.42^{\ddagger}$  & 3.36 \\
    
    DeepSeek-Coder-6.7B-Instruct & $3.23^{\ddagger}$ & $3.29^{\ddagger}$  & $3.32^{\ddagger}$  & $3.45^{\ddagger}$  & $3.48^{\ddagger}$  & 3.36 \\
    
    Mistral-7B-Instruct & $3.20^{\ddagger}$  & $3.27^{\ddagger}$  & $3.28^{\ddagger}$  & $3.42^{\ddagger}$  & $3.40^{\ddagger}$  & 3.31 \\
    
    CodeLlama-34B-Instruct & $3.11^{\ddagger}$  & $3.20^{\ddagger}$  & $3.21^{\ddagger}$  & $3.35^{\ddagger}$  & $3.23^{\ddagger}$  & 3.22 \\
    
    Llama-2-70B-Chat & $3.11^{\ddagger}$  & $3.22^{\ddagger}$  & $3.14^{\ddagger}$  & $3.38^{\ddagger}$  & $3.22^{\ddagger}$  & 3.21 \\ 
    
    WizardCoder-15B & $3.11^{\ddagger}$  & $3.03^{\ddagger}$  & $3.12^{\ddagger}$  & $3.27^{\ddagger}$  & $3.16^{\ddagger}$  & 3.14 \\
    
    CodeLlama-13B-Instruct & $3.05^{\ddagger}$  & $2.99^{\ddagger}$  & $3.15^{\ddagger}$  & $3.18^{\ddagger}$  & $3.25^{\ddagger}$  & 3.12 \\
    
    CodeLlama-7B-Instruct & $2.91^{\ddagger}$  & $3.11^{\ddagger}$  & $3.01^{\ddagger}$  & $3.18^{\ddagger}$  & $3.13^{\ddagger}$  & 3.07 \\
    
    WizardLM-33B & $3.07^{\ddagger}$  & $2.98^{\ddagger}$  & $2.91^{\ddagger}$  & $3.11^{\ddagger}$  & $3.10^{\ddagger}$  & 3.03 \\
    
    Llama-2-13B-Chat & $2.88^{\ddagger}$  & $2.97^{\ddagger}$  & $2.91^{\ddagger}$  & $3.18^{\ddagger}$  & $2.99^{\ddagger}$  & 2.98 \\
    
    WizardLM-7B & $2.63^{\ddagger}$  & $2.61^{\ddagger}$  & $2.51^{\ddagger}$  & $2.69^{\ddagger}$  & $2.64^{\ddagger}$  & 2.62 \\

    \arrayrulecolor{black}
    \bottomrule
    \end{tabular}
}
\end{table*}

In this section, we provide a detailed analysis of CodeUltraFeedback.
Specifically, we explore the inherent capabilities of LLMs to align with coding preferences. 
Furthermore, we showcase the diversity of the dataset in terms of LLM responses and scores. 

\subsection{LLMs Scores Exploration}

We start by exploring the ratings generated by GPT-3.5-Turbo in the LLM-as-a-Judge annotation phase. 
In Table~\ref{tab:initial_scores}, we present a detailed analysis of the average scores across coding preferences for each LLM, offering initial insight into their baseline capabilities.

At first glance, the scores might suggest marginal discrepancies given the 1--5 scoring range.
However, these variations are substantial and carry statistical significance.
To quantify this significance, we performed pairwise Welch's $t$-tests, comparing the score distribution for each LLM against GPT-3.5-Turbo across each coding preference.
This test was chosen because it accounts for differences in sample sizes and variances between groups.
For each comparison, the null hypothesis assumes that the means of the two groups are equal. 
We calculated p-values to assess the statistical significance of observed differences, using a significance threshold of $p < 0.05$.
The results reveal highly significant statistical differences, with $p < 0.001$, for most comparisons and validate the discernible performance gaps between all models and GPT-3.5-Turbo.
Moreover, the results also demonstrate a notably superior performance of GPT-4-Turbo relative to GPT-3.5-Turbo.

Overall, GPT-3.5-Turbo and GPT-4-Turbo score the highest across all preferences and on average.
This outcome is expected considering both models have undergone extensive instruction and reinforcement learning from human feedback (RLHF) tunings to align more closely with human preferences~\cite{ouyang2022training}.
The scores indicate that highly capable LLMs such as WizardCoder-33B and DeepSeek-Coder-33B-Instruct underperform compared to GPT-3.5-Turbo. 
This underperformance is likely due to a lack of alignment of the LLMs, and tuning these models using alignment techniques could improve their performance.

\subsection{LLMs Responses Analysis}

To gain more insights into the diversity of CodeUltraFeedback, we analyze the best and worst responses for each sample in terms of their assigned scores.

\begin{figure}[!t]
    \centering
    \includegraphics[width=.8\textwidth]{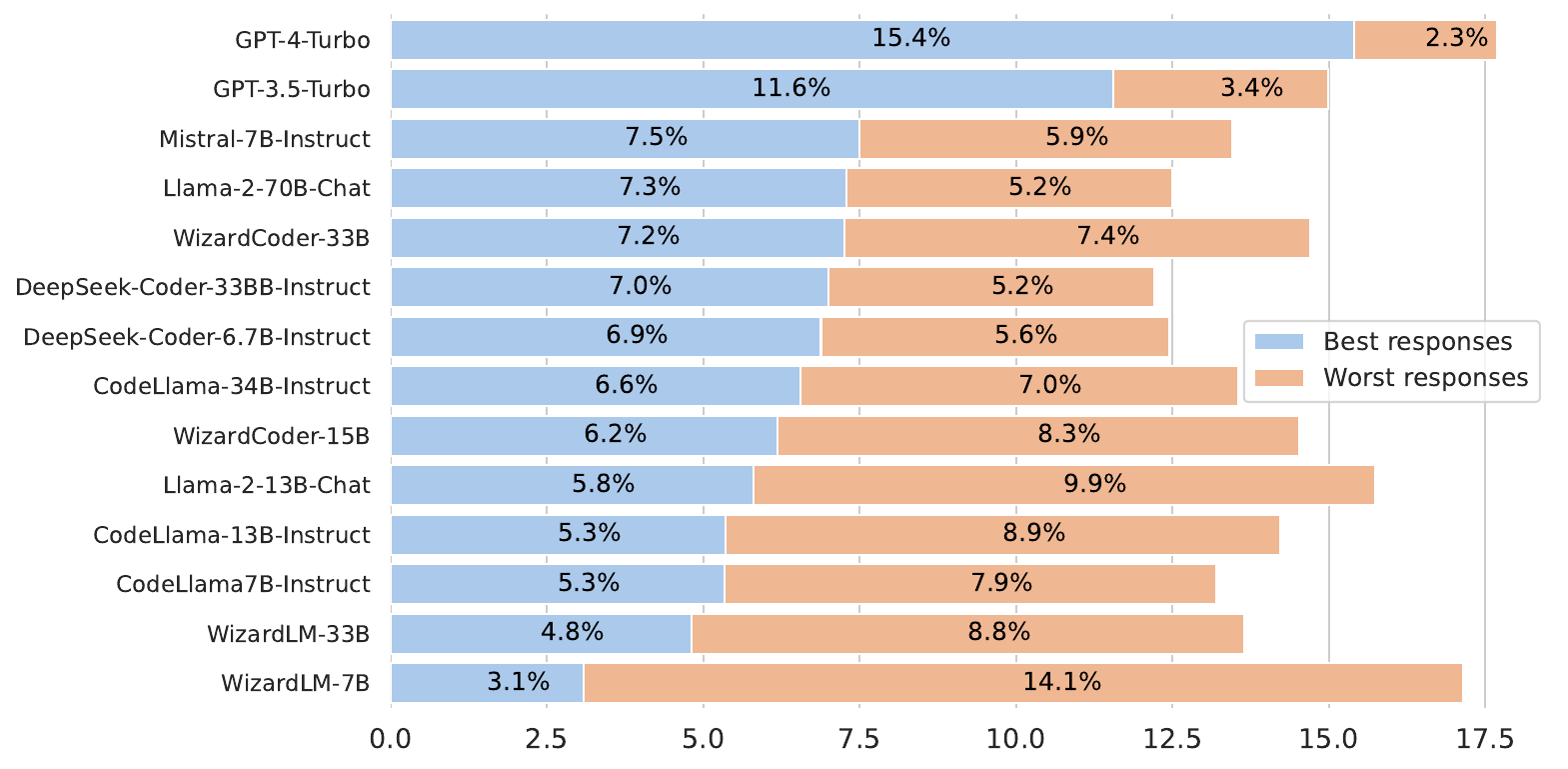}
    \vspace{-1em}
    \caption{Percentage of best and worst responses per LLM.}
    \label{fig:model_dist}
\end{figure}

Fig.~\ref{fig:model_dist} depicts the percentage of best and worst responses per LLM in the dataset. 
Although GPT-4-Turbo and GPT-3.5-Turbo consistently score the highest across all coding preferences on average, this plot highlights that another LLM response is favoured for 73\% of the samples.
Moreover, even LLMs with lower average scores produce the highest-rated responses for at least 3.1\% of the samples.
This property of our dataset is valuable for downstream applications.
For instance, it enables the alignment of LLMs using feedback data that encompasses a diverse range of models and mitigates the risk of the aligned LLM mimicking the behaviour of a single model.

\begin{figure}[!t]
    \centering
    \includegraphics[width=.6\textwidth]{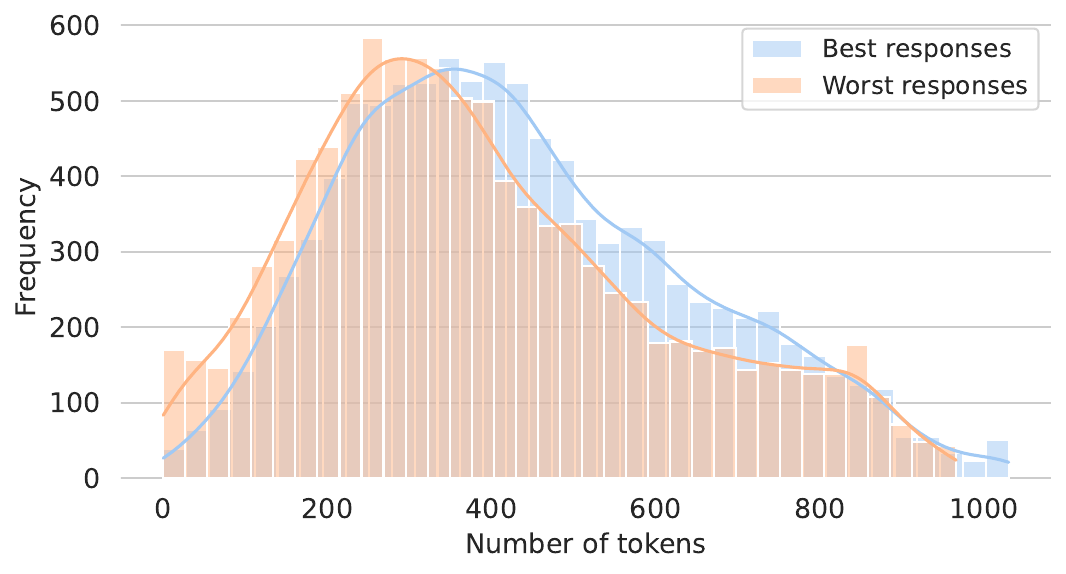}
    \vspace{-1em}
    \caption{Distribution of best and worst responses length.}
    \label{fig:len_dist}
\end{figure}s

Next, we analyze the distribution of best and worst response lengths in Fig.~\ref{fig:len_dist}.
Both distributions show similar patterns, though the best responses tend to be slightly longer.
Importantly, both distributions encompass a broad spectrum of response lengths.
This characteristic of CodeUltraFeedback is also valuable for model alignment, as it helps prevent the aligned LLM from developing a bias towards excessive verbosity.

\subsection{LLMs vs GPT-3.5-Turbo}

\begin{figure}[!t]
    \centering
    \includegraphics[width=.50\textwidth]{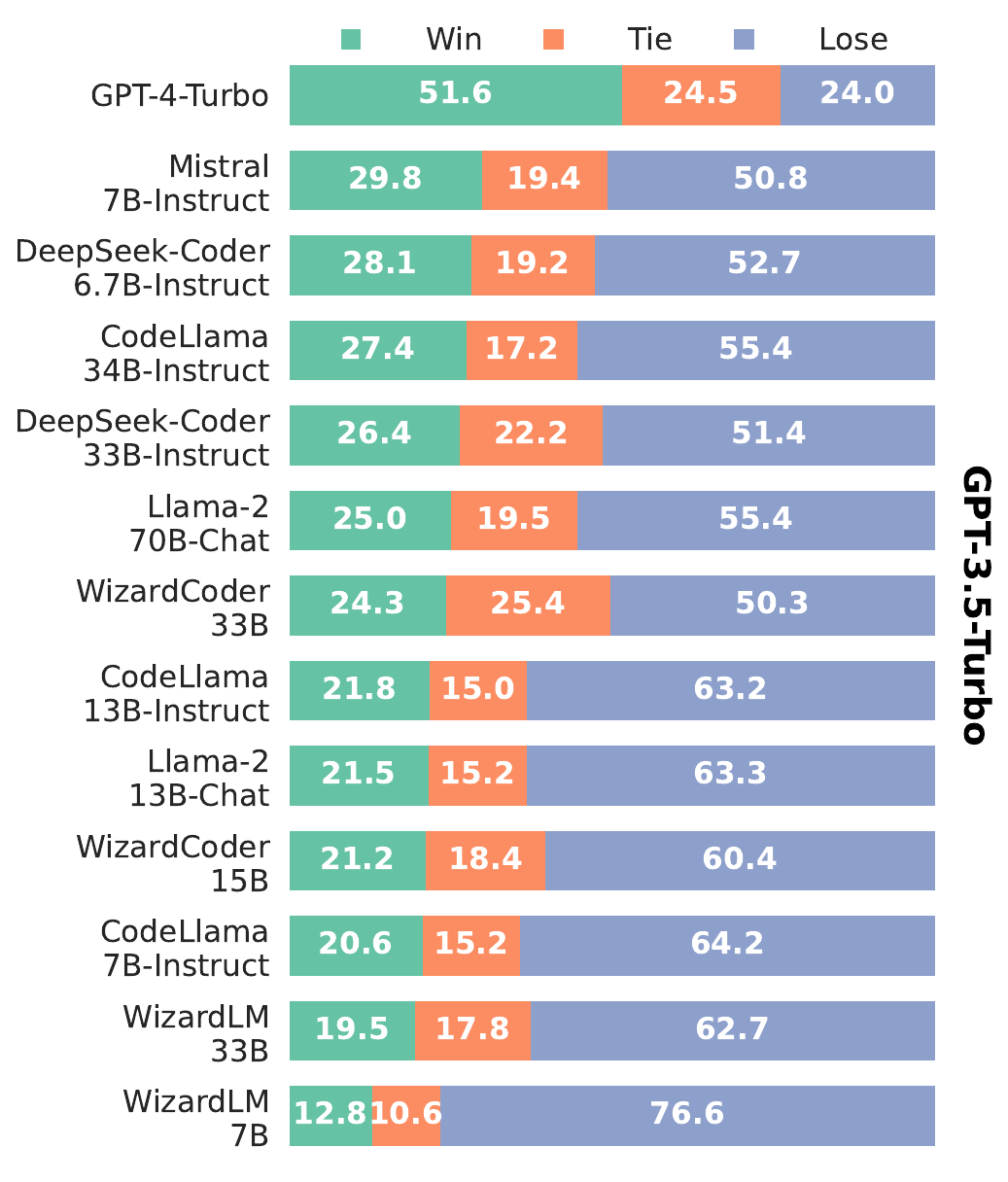}
    \vspace{-1em}
    \caption{Win-tie-lose ratios of LLMs against GPT-3.5-Turbo. An LLM wins/loses if it gets a greater/lower score than GPT-3.5-Turbo. A tie is when the LLM and GPT-3.5-Turbo get identical scores.}
    \label{fig:wtl}
\end{figure}

To gauge the relative performance of the LLMs against GPT-3.5-Turbo more in-depth, we analyze their performances through a win-tie-lose plot depicted in Fig.~\ref{fig:wtl}.
The idea is to pit each LLM (left-hand side of the figure, e.g., Mistral-7B-Instruct) against GPT-3.5-Turbo in a comparative match-up, where the model achieving the highest rating wins.
The plot illustrates the percentage of ratings of an LLM that are higher/lower compared to those of GPT-3.5-Turbo, all preferences combined (a tie stands for ratings having identical values).
This figure further demonstrates significant gaps between LLMs and GPT-3.5-Turbo, underscoring a consistent preference for GPT-3.5-Turbo and GPT-4-Turbo's responses over other LLMs.
For instance, even Mistral-7B-Instruct, the second-best LLM, achieves a win rate of merely 29.8\%. 
Additionally, the plot further validates the utilization of GPT-3.5-Turbo as a judge, showcasing its capability to discern between responses of differing quality.
For example, it is proficient in recognizing higher-quality responses, as evidenced by its comparison with GPT-4-Turbo, which boasts a win rate of 51.6\%.

In conclusion, these initial observations highlight a significant research opportunity and the necessity to delve deeper into aligning LLMs with coding preferences to make them more competitive with models like GPT-3.5 and GPT-4.

\section{Aligning LLMs to Coding Preferences}
\label{sec:aligning}

In this section, we present our experimental setup to improve LLM alignment with coding preferences using CodeUltraFeedback as preference data, which features the utilization of SFT and DPO~\cite{liu2023your, tunstall2023zephyr}.
The purpose is twofold: validate the utility of CodeUltraFeedback for LLM alignment to coding preferences and show that a small LLM tuned using SFT and DPO can achieve greater alignment performance.

\subsection{Supervised Fine-Tuning (SFT)}

As demonstrated in Zephyr's paper~\cite{tunstall2023zephyr}, an instruction-tuning phase is required prior to tuning an LLM using DPO as it facilitates tuning. 
This phase is achieved through supervised fine-tuning (SFT) using a dataset $\mathcal{D} = \{(x_1, y_1), ..., (x_n, y_n)\}$ of instruction-response pairs. 
Recent work leverages the Self-Instruct~\cite{wang2022self} and Evol-Instruct~\cite{xu2023wizardlm, luo2023wizardcoder} frameworks to compose such a dataset by generating instructions and responses using a highly capable model such as GPT-3.5/4~\cite{ouyang2022training, taori2023alpaca, wei2023magicoder}.
In this context, SFT is a form of knowledge distillation~\cite{xu2024survey} that leverages a supervised signal from responses generated by a teacher LLM to tune a student LLM. 

Formally, for each instruction-response pair $(x_i, y_i) \in \mathcal{D}$, the learning objective is to minimize the cross-entropy loss $\mathcal{L}_{SFT}$, defined as:
$$
\mathcal{L}_{SFT} = -\frac{1}{n}\sum_{i=1}^{n} \log P(y_i | x_i, \theta) ,
$$
where $P(y_i | x_i, \theta)$ represents the probability of generating the response $y_i$ given the instruction $x_i$, parameterized by $\theta$, the LLM parameters.

\subsection{Direct Preference Optimization (DPO)}

DPO is an efficient ranking optimization method to align LLMs to preference data, where a response $y_w$ is preferred over a rejected response $y_l$. 
Unlike traditional reinforcement learning methods like RLHF, which require training a separate reward model, DPO enhances stability and performance by directly adjusting the LLM's policy towards higher-ranked responses (i.e., $y_w$).
Given a dataset $\mathcal{P}$ of triplets $(x, y_w, y_l)$, a reference LLM policy $\pi_{ref}$, and an LLM policy $\pi_\theta$, the objective is to maximize the following expectation:
$$
\mathbb{E}_{\substack{(x, y_w, y_l)\sim \mathcal{P}}} \log \sigma \left( \beta \log \left( \frac{\pi_\theta(y_w | x)}{\pi_{ref}(y_w | x)} - \frac{\pi_\theta(y_l | x)}{\pi_{ref}(y_l | x)} \right) \right) .
$$
The expression $\left( \frac{\pi_\theta(y_w | x)}{\pi_{ref}(y_w | x)} - \frac{\pi_\theta(y_l | x)}{\pi_{ref}(y_l | x)} \right)$ calculates the difference in probabilities that the model assigns to the preferred response $y_w$ and the response $y_l$, relative to the reference policy $\pi_{ref}$.
Intuitively, the difference indicates how much the preference alignment of $\pi_\theta$ has improved for a given triplet $(x, y_w, y_l)$ after applying DPO compared to the reference policy $\pi_{ref}$.
$\beta$ is a hyperparameter that adjusts the sensitivity to the reference policy $\pi_{ref}$.
In our experiments, $\pi_{ref}$ is a SFT-tuned LLM.

\subsection{Experimental Details}
\label{sec:exp_details}

\noindent \textit{Datasets.} We build CodeUltraFeedback-Bench, a new benchmark for assessing LLM alignment to coding preferences.
CodeUltraFeedback-Bench consists of 500 randomly selected samples from CodeUltraFeedback, with a balanced representation of each preference.
The benchmark aims to provide a rigorous and comprehensive framework to evaluate and compare LLM alignment to coding preferences, allowing researchers to evaluate the impact of new alignment methods.

For SFT, we use Magicoder Evol-Instruct~\cite{wei2023magicoder}, which consists of instruction-response pairs generated by GPT-4.
We filter out the 10,000 samples used to build CodeUltraFeedback to mitigate data leakage between the SFT and DPO phases and avoid potential overfitting.
The process results in 100,772 samples (95\% for training, 5\% for model evaluation).

We use the remaining 9,500 samples from CodeUltraFeedback for DPO. 
From each sample, we extract binary preferences ($y_w$, $y_l$), denoting the preferred and rejected response.
These preferences are selected based on the highest and lowest ratings assigned by GPT-3.5-Turbo during the LLM-as-a-Judge annotation phase (see Section~\ref{sec:llm_annotation}).
In instances where multiple responses have the highest/lowest ratings, we select one randomly to obtain $y_w$\:/\:$y_l$.

\vspace{0.5em}

\noindent \textit{LLMs.} We conduct our experiments on CodeLlama-7B-Instruct as it comprises 7B parameters, allowing for tuning on a modest computing infrastructure. 
Furthermore, the model lags behind more proficient LLMs in CodeUltraFeedback's scores (see Table~\ref{tab:initial_scores}), which makes it an ideal candidate for demonstrating the potential impact of SFT and DPO on improving LLM alignment.
For comparison, we include the following LLMs: CodeLlama-13/34B-Instruct, DeepSeek-Coder-6.7B-Instruct, WizardCoder-33B, GPT-3.5-Turbo, and GPT-4-Turbo.

\vspace{0.5em}

\noindent \textit{Grading Procedure.} 
CodeUltraFeedback-Bench seeks to systematically compare LLM alignment, whereas CodeUltraFeedback's main purpose is to serve as a training dataset for preference tuning.
Therefore, we designed a new grading procedure adapted for evaluating each LLM individually using LLM-as-a-Judge.
We implement a reference-guided single-answer grading system that has proven efficient in alignment benchmarks such as MT-Bench~\cite{zheng2023judging}.
Each LLM's response is individually evaluated against a reference response, providing a consistent basis for comparison and ensuring objective judgments across LLMs.
Additionally, we instruct the judge LLM to provide a rationale alongside the rating, enhancing the consistency of the evaluations. 
Ratings are assigned on a scale from 1 to 10, facilitating nuanced assessments of LLM alignment.
We generate reference responses using GPT-3.5-Turbo and GPT-4-Turbo and use both models as judges.

\vspace{0.5em}

\noindent \textit{Training Details.} We use HuggingFace TRL~\cite{trl} to implement SFT and DPO with QLoRA~\cite{dettmers2023qlora} for efficient fine-tuning.
We use a cosine learning rate scheduler with a learning rate of 2e-04 for SFT and 5e-05 for DPO, and 10\% warmup steps. 
We fine-tune CodeLlama-7B-Instruct for 3 and 5 epochs for SFT and DPO, respectively.
All experiments were conducted on a single NVIDIA RTX A5000 GPU (24GB).
We provide extensive training details in our replication package.

\section{Results}

\begin{figure*}[!t]
    \centering
    \includegraphics[width=.85\linewidth]{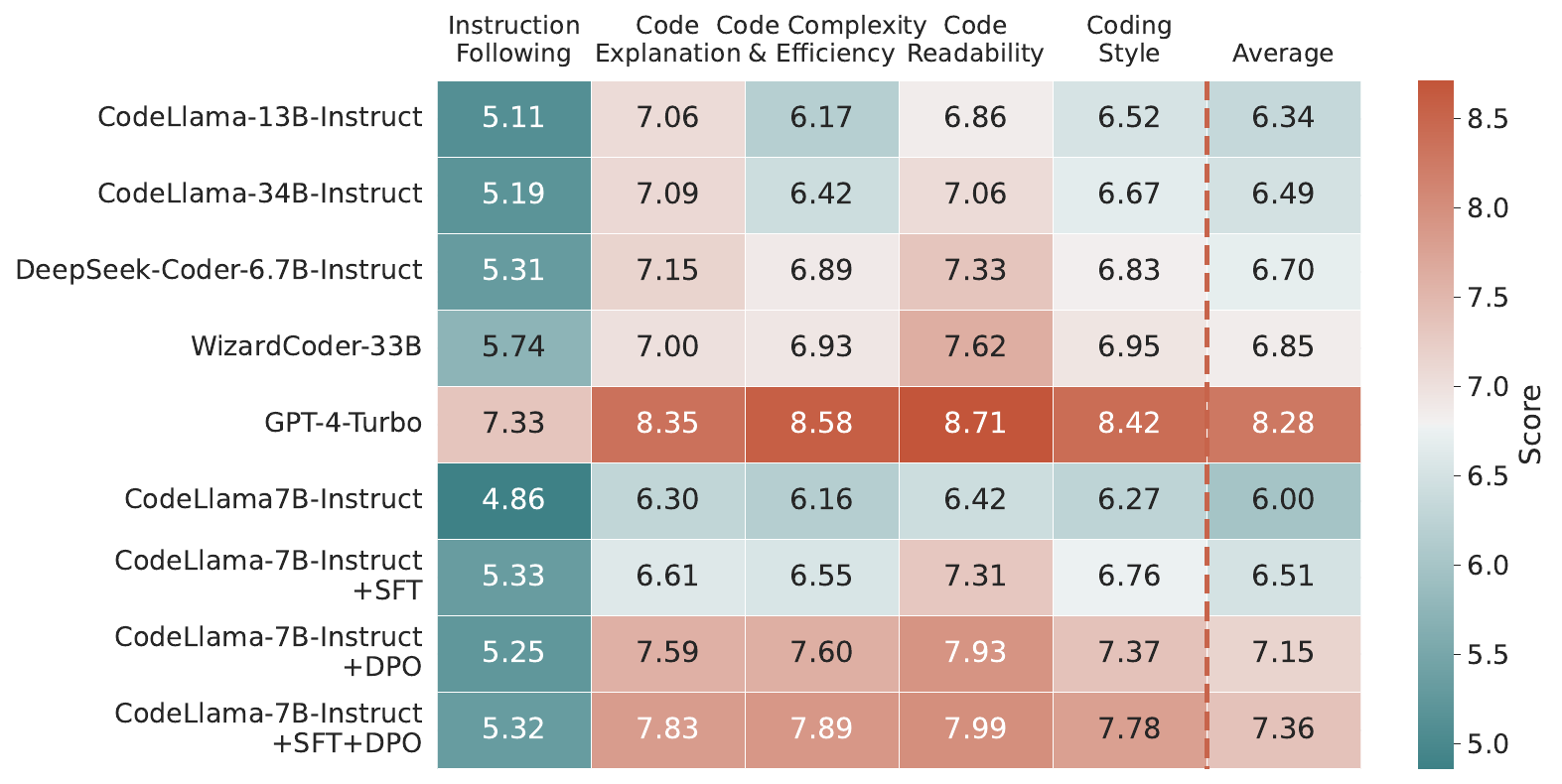}
    \vspace{-1em}
    \caption{Average alignment scores for LLMs across coding preferences on CodeUltraFeedback-Bench, evaluated using GPT-3.5-Turbo as a judge with reference-guided single-answer grading.}
    \label{fig:heatmap}
\end{figure*}

\noindent \textbf{SFT and DPO Improve Alignment to Coding Preferences.}
In Fig.~\ref{fig:heatmap}, we report the average alignment scores of the LLMs on the five coding preferences using GPT-3.5-Turbo as a judge with its own responses to the instructions as references.
We do not include GPT-3.5-Turbo in the results as references and responses would be identical, leading to biased and perfect alignment scores.

GPT-4-Turbo stands out by surpassing all other LLMs across every coding preference by a considerable margin. 
Interestingly, all LLMs demonstrate lower scores on the instruction following preference.
This trend suggests a relative challenge in achieving high performance in this area, possibly due to the inherently less precise nature of the preference and potentially vague instructions.
Conversely, LLMs tend to achieve higher scores on code explanation and readability preferences.
We hypothesize this trend can be attributed to the fact that all LLMs have undergone instruction tuning, which can encourage LLMs to output reasoning and explanations alongside generated code.
We observe that scores substantially increase when tuning CodeLlama-7B-Instruct with SFT+DPO, highlighting the effectiveness of these methods in enhancing model alignment across all coding preferences.
For instance, the model yields a $7.89$ score on code complexity and efficiency preference, substantially surpassing CodeLlama-7B-Instruct ($6.16$) and larger models, including Wizard-33B ($6.93$) and CodeLlama-34B-Instruct ($6.42$). 

\begin{figure}[!t]
    \centering
    \includegraphics[width=\linewidth]{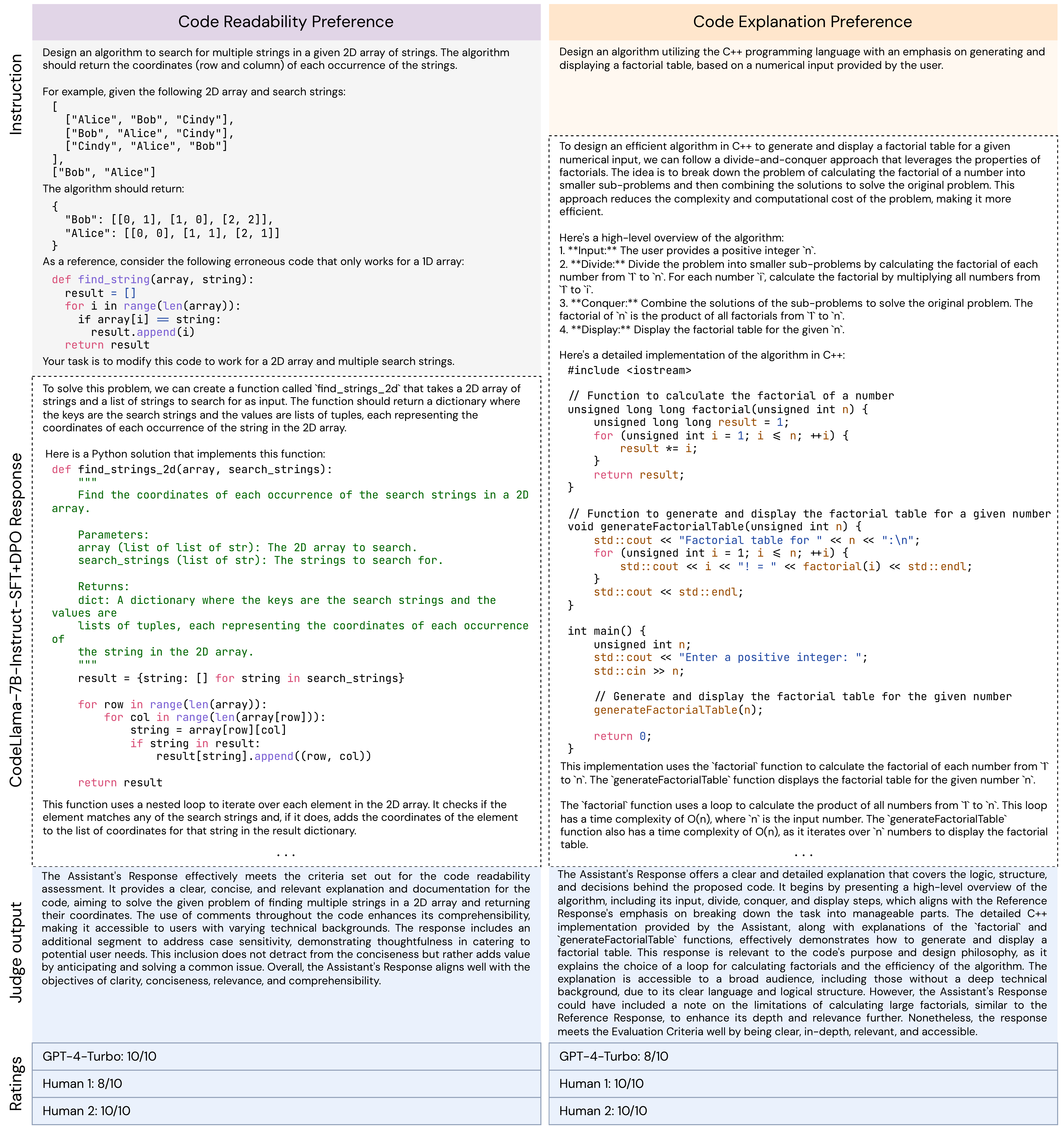}
    \caption{Examples of solutions generated by CodeLlama-7B-Instruct+SFT+DPO for two coding preferences: code readability and code explanation. }
    \label{fig:examples}
\end{figure}

Tuning CodeLlama-7B-Instruct with SFT shows an improvement in alignment scores compared to the base model, which are further elevated with the application of DPO independently or in conjunction with SFT.
To quantify these observations, we conducted pairwise $t$-tests (not reported for brevety) comparing the baseline model, against its variants tuned with SFT, DPO, and SFT+DPO.
Although SFT alone improves performance, we found no significant statistical difference with the baseline across preferences, except for code readability.
However, we found highly significant statistical differences ($p < 0.001$) for DPO and SFT+DPO for all preferences with the exception of instruction following.
After DPO and SFT+DPO tunings, CodeLlama-7B-Instruct achieves superior alignment compared to substantially larger models such as CodeLlama-34B-Instruct and WizardCoder-33B across all coding preferences.

To further illustrate the alignment of CodeLlama-7B-Instruct+SFT+DPO, we highlight two examples of solutions in Fig.~\ref{fig:examples} for the code readability and explanation preferences. 
For the code readability problem, the model's response demonstrates clear organization, use of comments, and adherence to the task requirements, earning high ratings (GPT-4: 10/10, Human-1: 8/10, Human-2: 10/10).
Similarly, in the code explanation problem, the model provides a well-structured, detailed breakdown of the algorithm, including logical flow and computational complexity, achieving a perfect score from all evaluators (GPT-4: 10/10, Human-1: 10/10, Human-2: 10/10).
These examples highlight the high-quality outputs of CodeLlama-7B-Instruct+SFT+DPO and its ability to generate solutions tailored to different coding preferences.

\begin{figure}[!ht]
    \centering
    \includegraphics[width=.6\linewidth]{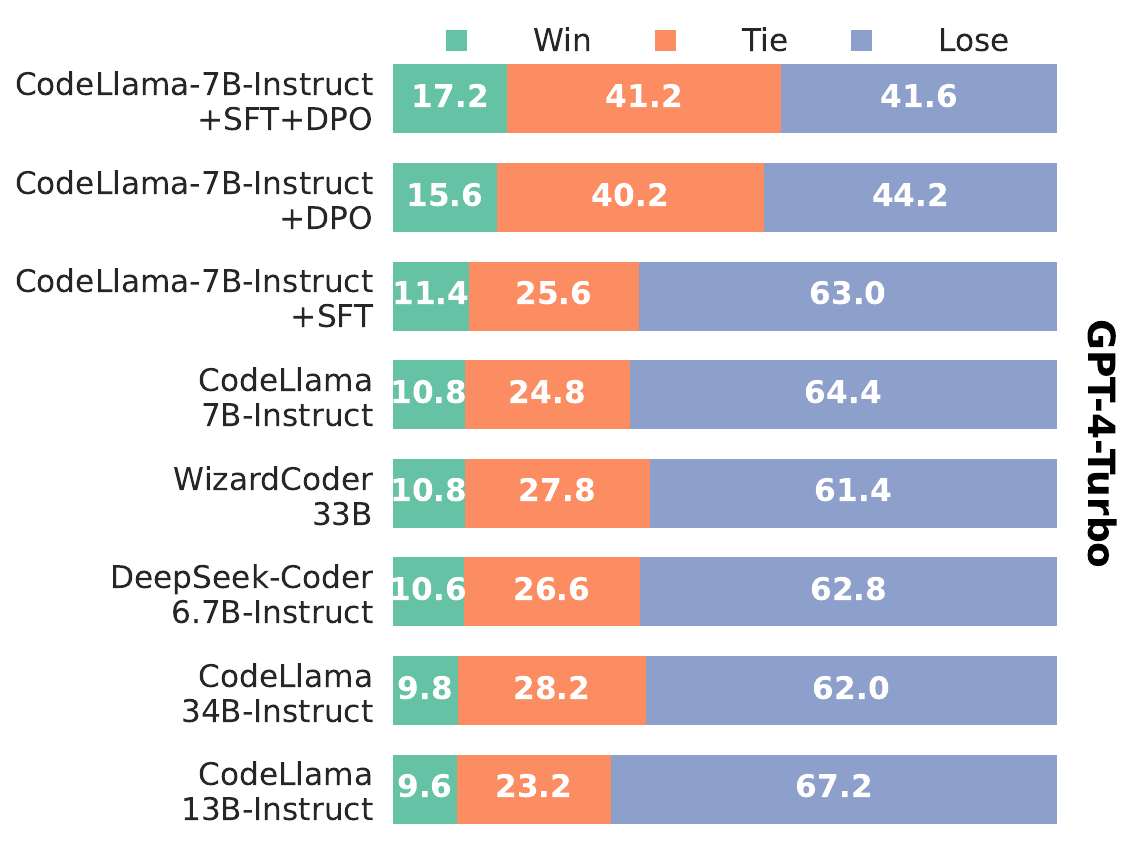}
    \vspace{-1em}
    \caption{Win-tie-lose ratios of LLMs against GPT-4-Turbo. DPO and SFT+DPO substantially increase the percentage of won and tie matches.}
    \label{fig:bench-wtl}
\end{figure}

\vspace{0.5em}
\noindent \textbf{DPO Enhances Competitive Edge Against GPT-4-Turbo.} Fig.~\ref{fig:bench-wtl} illustrates the win-tie-lose ratios of the LLMs against GPT-4-Turbo for all preferences combined.

Non-aligned models achieve low win rates, with CodeLlama-7B-Instruct and WizardCoder-33B peaking at a 10.8\% win rate.
GPT-4-Turbo, on the other hand, consistently outperforms most LLMs with win rates exceeding 60\%, underscoring its superiority. 
Tuning CodeLlama-7B-Instruct with DPO and SFT+DPO results in increases in win rates by 4.8\% and 6.2\%, respectively, and raises tie scores by 15.4\% and 16.4\%, respectively.
These improvements underscore the effectiveness of alignment techniques in rendering LLM responses more competitive.
Nonetheless, the remaining discernible gaps with GPT-4-Turbo suggest that there is still substantial room for improvement.

\vspace{0.5em}

\noindent \textbf{Different Judges and References Impact Alignment Scores.}
We investigate the effect of judges and references on the LLMs alignment scores in Table~\ref{tab:avg_llms_scores}.

Using GPT-3.5-Turbo as a judge, alignment scores are compared against its own (left values) and GPT-4-Turbo's responses (right values), revealing a decrease against the latter due to GPT-4-Turbo's higher response standards.
Additionally, CodeLlama-7B-Instruct+SFT+DPO achieves an alignment score of 7.08, nearing GPT-3.5-Turbo's average of 7.18.

Interestingly, divergent trends emerge under GPT-4-Turbo's judgment, with all CodeLlama models scoring lower, while other LLMs see score increases. 
For instance, CodeLlama-13B-Instruct's score falls from 5.58 to 4.83, while WizardCoder-33B's increases from 6.26 to 6.75.
Despite this, SFT, DPO, and SFT+DPO alignment techniques boost CodeLlama-7B-Instruct's performance, still exceeding the 34B variant of CodeLlama, albeit with varying impact with DPO.

These observations hint at GPT-4-Turbo's potential preference for certain response styles and variability in judgment compared to GPT-3.5-Turbo, underscoring the need for further investigation to fully understand the implications of judge selection on LLM alignment scores.

\renewcommand{\arraystretch}{1}
\setlength{\arrayrulewidth}{.5pt}
\begin{table}[!t]
\centering
\small
\caption{Average alignment scores of LLMs on CodeUltraFeedback-Bench. G-3.5 and G-4 refer to utilizing GPT-3.5-Turbo and GPT-4-Turbo responses as references, respectively.} 
\vspace{-1em}
\label{tab:avg_llms_scores}
    \begin{tabularx}{.65\linewidth}{l|@{}*{2}{C}@{}c}
    \toprule
    & \multicolumn{3}{c}{\textbf{Judge}} \\
    & \multicolumn{2}{c}{\textsc{GPT-3.5-Turbo}} & \textsc{GPT-4-Turbo} \\
    \hspace{8em} \textbf{Reference} & G-3.5 & G-4 & G-4 \\
    
    \midrule
    CodeLlama-13B-Instruct & 6.34 & 5.58 & 4.83 \\
    CodeLlama-34B-Instruct & 6.49 & 5.84 &  5.36 \\
    DeepSeek-Coder-6.7B-Instruct & 6.70 &  6.07 & 6.32 \\
    WizardCoder-33B & 6.85 & 6.26 &  \underline{6.75} \\
    GPT-3.5-Turbo & - & \textbf{7.18} &\textbf{7.35} \\
    GPT-4-Turbo & \textbf{8.28} & - &  -  \\
    
    \arrayrulecolor{black!50}
    \midrule
    
    CodeLlama-7B-Instruct & 6.00 &  5.46 &  4.72 \\
    \hspace{1em}+SFT & 6.51 &  5.83 & 5.84 \\
    \hspace{1em}+DPO &  7.15 &  6.79 &  5.08 \\
    \hspace{1em}+SFT+DPO &  \underline{7.36} & \underline{7.08} &  5.85 \\
    
    \arrayrulecolor{black}
    \bottomrule
    \end{tabularx}
\end{table}

\begin{figure}[!t]
    \centering

\begin{subfigure}[c]{0.45\linewidth}
    \centering
\small
\begin{tabular}{ll|cccc}
    \toprule
    \textbf{Preference} & & G-3.5 & G-4 & H-1 & H-2 \\
    
    \midrule
    \arrayrulecolor{black!50}
    
    \multirow{4}{*}{Instruction-Following} & G-3.5 & - & 0.41 & 0.29 & 0.32 \\
    & G-4 & - & - & 0.39 & 0.43 \\
    & H-1 & - & - & - &  0.57 \\
    & H-2 & - & - & - & - \\
    \midrule

    \multirow{4}{*}{Explanation} & G-3.5 & - & 0.60 & 0.78 & 0.67 \\
    & G-4 & - & - & 0.49 & 0.62 \\
    & H-1 & - & - & - & 0.64 \\
    & H-2 & - & - & - & - \\
    \midrule
    
    \multirow{4}{*}{Efficiency} & G-3.5 & - & 0.73 & 0.50 & 0.58 \\
    & G-4 & - & - & 0.49 & 0.56 \\
    & H-1 & - & - & - & 0.52 \\
    & H-2 & - & - & - & - \\
    \midrule

    \multirow{4}{*}{Readability} & G-3.5 & - & 0.53 & 0.83 & 0.91 \\
    & G-4 & - & - & 0.48 & 0.43 \\
    & H-1 & - & - & - & 0.64 \\
    & H-2 & - & - & - & - \\
    \midrule

    \multirow{4}{*}{Style} & G-3.5 & - & 0.48 & 0.77 & 0.67 \\
    & G-4 & - & - & 0.48 & 0.37 \\
    & H-1 & - & - & - & 0.60 \\
    & H-2 & - & - & - & - \\

    \arrayrulecolor{black}
    \bottomrule
    \end{tabular}
    \caption{Agreement between humans and LLMs judgements in terms of \emph{Cohen's kappa} coefficient with a tolerance threshold of $\tau=2$.}
    \label{fig:tab_agreements}
\end{subfigure}
\hfill
\begin{subfigure}[c]{0.49\linewidth}
    \centering
    \includegraphics[width=\linewidth]{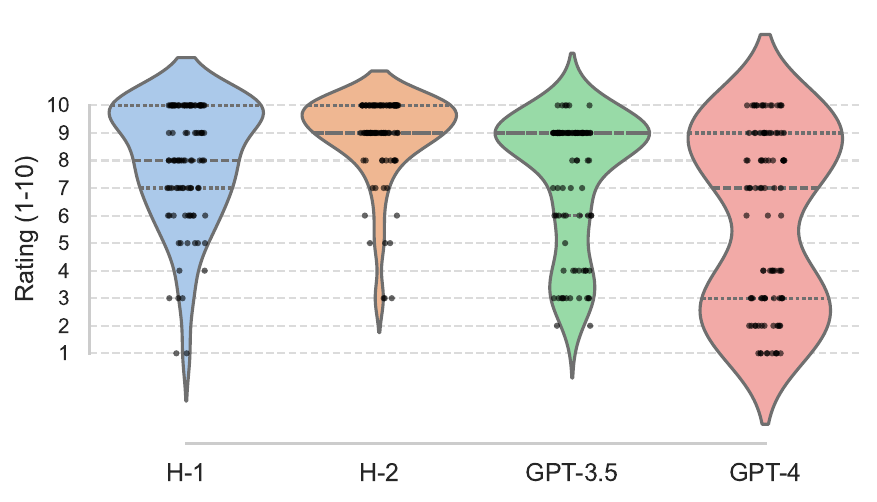}
    \caption{Score distribution of humans (H-1 and H-2) and GPT-3.5/4.}
    \label{fig:score_distribution}
\end{subfigure}
\vspace{-.5em}
\caption{Analysis of the judging agreements between humans and LLMs (a) and their score distributions (b) for solutions generated by CodeLlama-7B-Instruct+SFT+DPO.}
\label{fig:human_llm_scores}
\end{figure}

\vspace{0.5em}
\noindent \textbf{LLM Judgements Correlate with Human Evaluators.}
We investigate the judging capabilities of GPT-3.5 and GPT-4 by evaluating their alignment with human evaluators.
We randomly select 20 samples per coding preference, manually annotated by two authors of the paper following the same judging setup as LLM-as-a-Judge. 
Fig.~\ref{fig:tab_agreements} reports \emph{Cohen's kappa} coefficients ($\kappa$) for each LLM-human combination. 
Given that ratings are on a 1-10 scale, we use a tolerance threshold of $\tau=2$, considering ratings like 7 and 9 to be in agreement.
We also compare the score distributions of humans and LLMs in Fig.~\ref{fig:score_distribution}.

We observe that GPT-3.5 generally shows stronger agreement with human evaluators (H-1 and H-2) across coding preferences compared to GPT-4. 
For example, GPT-3.5 demonstrates almost perfect agreement with both humans in code readability ($\kappa=0.83$ and $\kappa=0.91$), whereas GPT-4 demonstrates only moderate agreement in this preference ($\kappa=0.48$ and $\kappa=0.43$). 
Interestingly, for most coding preferences—except for instruction following—GPT-3.5's judgments show stronger alignment with individual human judges than the agreement observed between the human judges themselves.
This observation suggests that GPT-3.5's evaluations may serve as a reliable proxy for human judgment in these coding preferences.

The score distributions in Fig.~\ref{fig:score_distribution} reveal that humans tend to give higher ratings compared to GPT-3.5 and GPT-4.
Notably, GPT-3.5 shows better consistency with human ratings', while GPT-4's ratings are more evenly distributed, which contributes to its lower kappa values.
Overall, our findings highlight that GPT-3.5 aligns more closely with human evaluators, though both LLMs would benefit from further calibration, particularly in the instruction-following preference.

\renewcommand{\arraystretch}{1}
\setlength{\arrayrulewidth}{.5pt}

\vspace{0.5em}

\noindent \textbf{SFT and DPO Improve Functional Correctness.} 
While our findings highlight the positive impact of SFT and DPO on aligning LLMs with coding preferences, ensuring the functional correctness of the generated code remains a pivotal concern.
In Table~\ref{tab:correctness}, we report CodeLlama-7B-Instruct's Pass@$k$ on HumanEval~\cite{chen2021evaluating} and HumanEval+~\cite{liu2023your}.

Firstly, SFT substantially enhances Pass@$k$ on both benchmarks, with Pass@$1$ rising from 37.9 to 51.2 and 33.2 to 45.6 for HumanEval and HumanEval+, respectively.
This result aligns with Magicoder~\cite{wei2023magicoder} and WizardCoder~\cite{luo2023wizardcoder} findings, showing that knowledge distillation through SFT substantially enhances the LLM's effectiveness on these benchmarks.
Secondly, while DPO and SFT+DPO variants exhibit a slight reduction in Pass@$k$ relative to SFT alone, they still maintain substantially higher Pass@$k$ compared to the baseline. 

These results show that LLM alignment through SFT and DPO also improves functional correctness, mostly due to the SFT phase, further demonstrating the usefulness of aligning LLMs.
The lesser improvements in Pass@$k$ with DPO can be attributed to DPO's learning objective, focusing on favouring preferred responses with respect to coding preferences rather than based on their functional correctness. 
Therefore, we underscore the need for more investigations into the design of learning objectives and methods that can prioritize both functional correctness and alignment to coding preferences.

\renewcommand{\arraystretch}{1}
\setlength{\arrayrulewidth}{.5pt}
\begin{table}[!t]
\centering
\caption{Pass@$k$ of CodeLlama-7B-Instruct variants on HumanEval and HumanEval+.} 
\vspace{-1em}
\label{tab:correctness}
\resizebox{.5\linewidth}{!}{%
    \begin{tabular}{l|cccc}
    \toprule
    & \multicolumn{2}{c}{\textbf{HumanEval}} & \multicolumn{2}{c}{\textbf{HumanEval+}} \\
    & $k$=1 & $k$=10 & $k$=1 & $k$=10 \\
    
    \midrule
    
    CodeLlama-7B-Instruct & 37.9 & 60.4 & 33.2 & 54.9 \\
    \hspace{1em}+SFT & \textbf{51.2} & \textbf{82.9} & \textbf{45.6} & \textbf{79.3} \\
    \hspace{1em}+DPO & 42.3 & \underline{80.5} & 35.8 & \underline{70.1} \\
    \hspace{1em}+SFT+DPO & \underline{43.1} & 75.6 & \underline{36.7} & 69.5 \\
    
    \arrayrulecolor{black}
    \bottomrule
    \end{tabular}
}
\end{table}

\vspace{.5em}

\begin{figure}[!ht]
    \centering
    \begin{subfigure}[t]{0.48\textwidth}
        \centering
        \includegraphics[width=\linewidth]{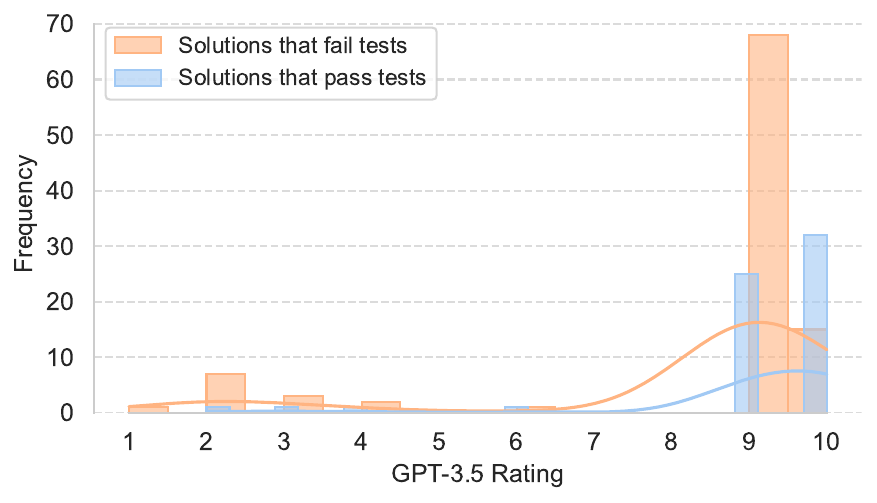}
    \end{subfigure}%
    ~ 
    \begin{subfigure}[t]{0.48\textwidth}
        \centering
        \includegraphics[width=\linewidth]{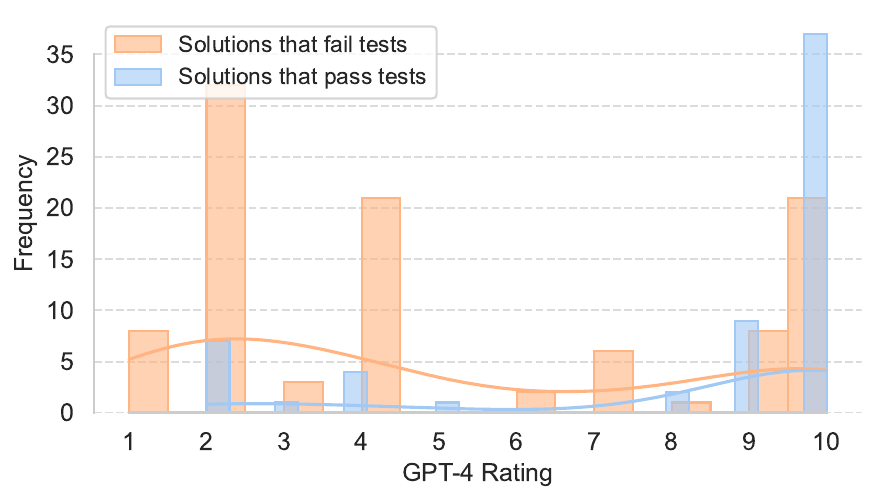}
    \end{subfigure}
    \vspace{-1em}
    \caption{GPT-3.5 and GPT-4 ratings for HumanEval+ solutions generated by CodeLlama-7B-Instruct+SFT+DPO.}
    \label{fig:gpt_ratings_humaneval}
\end{figure}

\noindent \textbf{GPT-4 Prefers Test-Passing Solutions.}
To further analyze the judging behavior of GPT-3.5 and GPT-4, we examine their ratings for solutions generated by CodeLlama-7B-Instruct+SFT+DPO on the HumanEval+ benchmark. 
Fig.~\ref{fig:gpt_ratings_humaneval} depicts the distribution of ratings assigned by GPT-3.5 (left) and GPT-4 (right), distinguishing between solutions that pass and fail the corresponding test cases.
To compute the ratings, we followed an identical LLM-as-a-Judge setup than for CodeUltraFeedback by designing a specific prompt (see Table~\ref{tab:humaneval_rating}), which focuses on judging functional correctness and the usefulness of the solution.

The results indicate a significant behavioral difference between the GPT-3.5 and GPT-4.
GPT-4 shows a clear preference for solutions that pass test cases, assigning them consistently higher ratings. 
In contrast, GPT-3.5 demonstrates little distinction between test-passing and test-failing solutions, as reflected in the more uniform rating distribution across both categories.
These findings underscore GPT-4's strength in aligning ratings with objective evaluation criteria, such as passing test cases, and highlight the need for further refinement of GPT-3.5 in this regard.

\section{Discussion}

In this paper, we have demonstrated how LLM-as-a-Judge enables the creation of new datasets and benchmarks like CodeUltraFeedback and CodeUltraFeedback-Bench to facilitate the tuning of LLMs for better alignment with coding preferences and assess their performance.
This section offers further insights from our experiments, potential improvements for LLM alignment and evaluation, and the significance of exploring various LLM-as-a-Judge configurations.
Additionally, we outline some limitations of CodeUltraFeedback and CodeUltraFeedback-Bench.

\subsection{Limitations of Existing Evaluation Metrics}
Evaluation metrics are central to assessing the performance of LLMs for code generation. 
Automated metrics and static analysis tools provide scalable and objective evaluations. 
However, while these techniques effectively capture specific aspects of the LLM-generated code, e.g., whether it passes unit tests or matches a ground truth solution, they fail to evaluate the latter against user-centric preferences such as code readability or whether the response strictly follows the user's instructions. 

One of the main limitations of automated metrics is that they cannot assess language subtleties as humans do. 
For instance, metrics like Pass@$k$ are useful for evaluating functional correctness but cannot discriminate between an overly verbose solution or one lacking meaningful comments and explanations. 
On the other hand, static analysis tools focus on enforcing rules, which does not encompass any form of language understanding that would allow determining whether a solution includes useful explanations for a given generated code. 
Altogether, our approach featuring LLM-as-a-Judge offers a solution to these limitations by focusing on LLM alignment with user coding preferences. 
For example, when evaluating a solution to a coding task, the judge LLM can discern whether comments add meaningful context, whether variable names enhance code readability, or whether an explanation is structured logically.
This process is parallel to how humans would judge a solution.

The interplay between LLM-as-a-Judge and automated metrics offers significant opportunities for advancing evaluation methodologies. 
In particular, the combination of automated evaluation metrics and judging LLMs could result in more holistic evaluations of LLM-generated code and enhance the way we benchmark code LLMs.
For instance, future metrics might integrate functional correctness scores with readability and explanation assessments, creating a weighted evaluation system that reflects the multifaceted nature of high-quality code. 

\subsection{LLM Alignment and LLM-as-a-Judge}

\noindent \textbf{Balancing Learning Objectives.} 
A crucial area requiring deeper investigation is the balance between functional correctness and non-functional requirements. 
In our experiments, we demonstrate that CodeLlama-7B-Instruct+SFT+DPO shows higher performance on HumanEval/HumanEval+ compared to the base model.
Understanding how to integrate functional and non-functional learning objectives could significantly advance the tuning process of LLMs and further enhance their performance across a broad range of benchmarks.

\vspace{.5em}

\noindent \textbf{CodeUltraFeedback for Critic LLM Training.}
In this work, we demonstrate the utility of CodeUltraFeedback for LLM preference tuning using RLAIF and DPO. 
One of the drawbacks of CodeUltraFeedback-Bench is the need to leverage models like GPT-4 to evaluate other LLMs, which might turn out to be cost-prohibitive. 
Nonetheless, we believe the AI feedback data in CodeUltraFeedback, e.g., ratings and rationales, can be leveraged to fine-tune a small critic LLM trained to evaluate other LLMs. 
Prior work including Shepherd~\cite{wang2023shepherd} and Prometheus~\cite{kim2023prometheus} have been proposed around that idea and showed promising results.

\vspace{.5em}
\noindent \textbf{Parallel Findings with Zephyr} 
Our findings draw parallels with Zephyr 's~\cite {tunstall2023zephyr} insights on DPO, with models beginning to overfit after a few epochs, achieving 100\% training set accuracy. 
In our case, overfitting after five DPO training epochs correlated with improved performance.
Although the model overfits, the gap between chosen and rejected rewards continued to grow, suggesting that monitoring reward accuracy might not be relevant for model selection.  
Therefore, more experimentation on DPO could illuminate better model tuning and selection approaches for improved model alignment.

\vspace{.5em}

\noindent \textbf{Potential Judgment Biases.} We highlight GPT-3.5-Turbo's robustness in judgment in both our dataset and benchmark. 
The performance ranking of LLMs remained consistent, underscoring the reliability of GPT-3.5-Turbo's evaluations.
Despite this, there remains the possibility of inherent biases in LLM judgments, such as a preference for lengthy or verbose responses.
The grading procedures employed for CodeUltraFeedback and CodeUltraFeedback-Bench were carefully designed, drawing on established practices to preclude such biases~\cite{zheng2023judging, li2023alpacaeval}.
Single-answer grading with a reference response to the instruction and chain-of-thought prompting enables both GPT-3.5-Turbo and GPT-4-Turbo to produce consistent judgements.
However, further exploration into the potential biases influencing LLM judgments is warranted.

\vspace{.5em}

\noindent \textbf{Exploring Alternative Judges.}
Future studies could benefit from incorporating a wider range of judges to understand how different LLMs' evaluative perspectives might influence the judgment process and outcomes.
Additionally, the reliance on closed-source models can be mitigated by tuning small and open-source critic LLMs.

\subsection{Limitations}

Our methodology presupposes the existence of a highly capable model, such as GPT-3.5 or GPT-4, as a proxy to humans to accurately evaluate other LLMs. 
The reliance on these models is predicated upon previous findings, which indicate that GPT-3.5 and GPT-4 align closely with human judgments~\cite{zheng2023judging, li2023alpacaeval}, achieving agreement rates on par with those between humans themselves on MT-Bench~\cite{zheng2023judging}.
Additionally, recent work~\cite{zhuo2024ice} demonstrates that ICE-Score, a novel metric for code generation based on LLM-as-a-Judge is superior to prior metrics and has a higher correlation with human judgements. 
Additionally, we performed a correlation analysis between GPT-3.5, GPT-4, and human evaluators using 100 samples from CodeUltraFeedback-Bench. 
The results reveal that GPT-3.5 shows a strong correlation with human judgments across coding preferences, while exhibiting a similar score distribution.
Such evidence supports the utilization of LLMs as evaluators.

Our current iteration of CodeUltraFeedback-Bench relies on randomly selected samples. 
Future versions aim to employ a more rigorous filtering process, possibly incorporating human annotations, to enhance the quality of selected samples. 

Lastly, our grading procedure in CodeUltraFeedback-Bench necessitates reference responses. 
This usage of reference responses gives the judge LLMs a consistent point of comparison, enabling fair and consistent judgements across LLMs outputs.
Future work might explore ways to alleviate this dependency.

\section{Related Work}

We differentiate our work from the existing landscape of datasets and benchmarks in automated software engineering, with a particular focus on execution-based benchmarks, closed-solutions datasets, and benchmarks, as well as those assessing non-functional aspects of generated code.

\subsection{Execution-based Benchmarks}

This category of benchmarks emphasizes the functional correctness of generated code by evaluating whether it passes a set of unit tests. 
Execution-based benchmarks include HumanEval~\cite{chen2021evaluating}, MBPP~\cite{austin2021program}, APPS~\cite{hendrycksapps2021}, and DS-1000~\cite{lai2023ds}, which have extensively been used to compare LLMs in code generation scenarios~\cite{roziere2023code, muennighoff2023octopack, luo2023wizardcoder, wei2023magicoder}. 
Subsequent work has expanded HumanEval and MBPP benchmarks to multiple programming languages~\cite{athiwaratkun2022multi, cassano2023multipl, zheng2023codegeex}, more unit tests~\cite{liu2023your}, and more tasks including code summarization~\cite{athiwaratkun2022multi}, code repair~\cite{muennighoff2023octopack}, code explanation~\cite{muennighoff2023octopack}, and code translation~\cite{zheng2023codegeex}.
This category also encompasses benchmarks designed around specific use cases, such as Odex~\cite{zhou2022docprompting}, which deals with Python open-domain problems and includes manual annotations of user intents, and StudentEval~\cite{babe2023studenteval}, which utilizes student-defined prompts for Python code generation. 
xCodeEval~\cite{khan2023xcodeeval} benchmark extends the scope to code understanding, generation, and retrieval across numerous tasks. 
Lastly, benchmarks like ClassEval~\cite{du2023classeval} and CoderEval~\cite{yu2024codereval} shift the focus towards evaluating LLMs' proficiency in generating functionally correct code across various programming abstractions, such as class, file, and project levels.

Our work differs from this category of benchmarks and focuses on aligning LLMs to coding preferences and measuring their alignment using an LLM as a proxy for human evaluation without relying on unit tests or automated metrics.
Nonetheless, we believe evaluating the functional correctness of LLM-generated code remains a pivotal concern complementary to the aim of CodeUltraFeedback and CodeUltraFeedback-Bench.

\subsection{General Datasets and Closed-Solutions Benchmarks}

Initial datasets like GitHub Java Corpus~\cite{allamanis2013mining}, Py150/Js150~\cite{raychev2016probabilistic, raychev2016learning}, and CodeSearchNet~\cite{husain2019codesearchnet} laid the groundwork for evaluating language models in code-related tasks, including modelling, summarization, and search. 
Subsequent developments introduced benchmarks like CodeXGlue~\cite{lu2021codexglue}, CodeNet~\cite{puri2021codenet}, XLCoST~\cite{zhu2022xlcost}, ReCode~\cite{wang2022recode}, and CrossCodeBench~\cite{niu2023crosscodebench}, each expanding the evaluation scope to include code understanding, generation, and robustness against perturbations across various coding tasks and languages. 
Recently, CrossCodeEval~\cite{ding2023crosscodeeval} broadened this scope by assessing generative capabilities and the use of cross-file information for project-wide coding tasks.

These datasets and benchmarks evaluate language models' core capabilities, mainly through task-specific transfer learning, relying on ground truth solutions which may overlook valid code variations. 
In contrast, CodeUltraFeedback and CodeUltraFeedback-Bench focus on aligning LLMs with human coding preferences. 
Additionally, the LLM-as-a-Judge approach serves as an alternative to automated evaluations, prioritizing nuanced assessment of natural and programming languages over strict adherence to ground truth.

\subsection{Non-Functional Evaluation of LLM-Generated Code}

Recent advancements have expanded LLM evaluation to include non-functional requirements~\cite{yang2024robustness}, an aspect overlooked in earlier studies. 
Examples include efforts to assess the quality of AI-generated code~\cite{liu2023refining, siddiq2023lightweight} by using static analysis tools.
In contrast, our work adopts a broader strategy, employing LLMs' advanced reasoning to tune and assess their alignment with human coding preferences.
Furthermore, Yetiştiren et al.~\cite{yeticstiren2023evaluating} evaluated LLM-generated code using quality metrics such as security, reliability, and maintainability on the HumanEval benchmark. 
Our work encompasses a broader evaluation scope, including more complex instructions, and CodeUltraFeedback to tune and align LLMs to coding preferences.
In contrast to CyberSecEval~\cite{bhatt2023purple} and EffiBench~\cite{huang2024effibench}, which concentrate on particular code aspects like security and efficiency, our methodology based on LLM-as-a-Judge provides a holistic evaluation across several dimensions, offering a framework that can be adapted to assess an array of coding preferences.
NoFunEval~\cite{singhal2024nofuneval} stands as the most closely related benchmark to our work, evaluating LLM-generated code's non-functional properties across 958 coding problems using functional specifications and static analysis tools. 
In contrast, we leverage LLMs' evaluative capabilities for LLM evaluation, overcoming static analysis tools' limitations in capturing language nuances.
Additionally, we introduce a comprehensive dataset for tuning LLM preferences, distinguishing our work as a unique contribution to the field of code LLM evaluation.

\section{Conclusion and Future Work}

In this paper, we introduce CodeUltraFeedback and CodeUltraFeedback-Bench, a preference dataset and benchmark of instructions for LLM alignment to five coding preferences. 
Our analysis reveals significant alignment disparities among various LLMs compared to GPT-3.5 and GPT-4.
Furthermore, we demonstrate how CodeUltraFeedback facilitates preference tuning using SFT, RLAIF and DPO.
Our experiments conclude that CodeLlama-7B-Instruct tuned with SFT and DPO outperforms 34B LLMs on CodeUltraFeedback-Bench and enhances functional correctness on HumanEval+.
We hope CodeUltraFeedback and CodeUltraFeedback-Bench can further support research related to LLM alignment in automated software engineering. 
In future work, we plan to explore more judges to assess LLMs on CodeUltraFeedback-Bench and use CodeUltraFeedback to fine-tune our own judge LLM, thereby reducing our dependence on closed-source models like GPT-4.


\bibliographystyle{ACM-Reference-Format}
\bibliography{references}


\begin{thebibliography}{67}


\ifx \showCODEN    \undefined \def \showCODEN     #1{\unskip}     \fi
\ifx \showDOI      \undefined \def \showDOI       #1{#1}\fi
\ifx \showISBNx    \undefined \def \showISBNx     #1{\unskip}     \fi
\ifx \showISBNxiii \undefined \def \showISBNxiii  #1{\unskip}     \fi
\ifx \showISSN     \undefined \def \showISSN      #1{\unskip}     \fi
\ifx \showLCCN     \undefined \def \showLCCN      #1{\unskip}     \fi
\ifx \shownote     \undefined \def \shownote      #1{#1}          \fi
\ifx \showarticletitle \undefined \def \showarticletitle #1{#1}   \fi
\ifx \showURL      \undefined \def \showURL       {\relax}        \fi
\providecommand\bibfield[2]{#2}
\providecommand\bibinfo[2]{#2}
\providecommand\natexlab[1]{#1}
\providecommand\showeprint[2][]{arXiv:#2}

\bibitem[Allamanis and Sutton(2013)]%
        {allamanis2013mining}
\bibfield{author}{\bibinfo{person}{Miltiadis Allamanis} {and} \bibinfo{person}{Charles Sutton}.} \bibinfo{year}{2013}\natexlab{}.
\newblock \showarticletitle{Mining source code repositories at massive scale using language modeling}. In \bibinfo{booktitle}{\emph{2013 10th working conference on mining software repositories (MSR)}}. IEEE, \bibinfo{pages}{207--216}.
\newblock


\bibitem[Athiwaratkun et~al\mbox{.}(2022)]%
        {athiwaratkun2022multi}
\bibfield{author}{\bibinfo{person}{Ben Athiwaratkun}, \bibinfo{person}{Sanjay~Krishna Gouda}, \bibinfo{person}{Zijian Wang}, \bibinfo{person}{Xiaopeng Li}, \bibinfo{person}{Yuchen Tian}, \bibinfo{person}{Ming Tan}, \bibinfo{person}{Wasi~Uddin Ahmad}, \bibinfo{person}{Shiqi Wang}, \bibinfo{person}{Qing Sun}, \bibinfo{person}{Mingyue Shang}, {et~al\mbox{.}}} \bibinfo{year}{2022}\natexlab{}.
\newblock \showarticletitle{Multi-lingual evaluation of code generation models}.
\newblock \bibinfo{journal}{\emph{arXiv preprint arXiv:2210.14868}} (\bibinfo{year}{2022}).
\newblock


\bibitem[Austin et~al\mbox{.}(2021)]%
        {austin2021program}
\bibfield{author}{\bibinfo{person}{Jacob Austin}, \bibinfo{person}{Augustus Odena}, \bibinfo{person}{Maxwell Nye}, \bibinfo{person}{Maarten Bosma}, \bibinfo{person}{Henryk Michalewski}, \bibinfo{person}{David Dohan}, \bibinfo{person}{Ellen Jiang}, \bibinfo{person}{Carrie Cai}, \bibinfo{person}{Michael Terry}, \bibinfo{person}{Quoc Le}, {et~al\mbox{.}}} \bibinfo{year}{2021}\natexlab{}.
\newblock \showarticletitle{Program Synthesis with Large Language Models}.
\newblock \bibinfo{journal}{\emph{arXiv preprint arXiv:2108.07732}} (\bibinfo{year}{2021}).
\newblock


\bibitem[Babe et~al\mbox{.}(2023)]%
        {babe2023studenteval}
\bibfield{author}{\bibinfo{person}{Hannah~McLean Babe}, \bibinfo{person}{Sydney Nguyen}, \bibinfo{person}{Yangtian Zi}, \bibinfo{person}{Arjun Guha}, \bibinfo{person}{Molly~Q Feldman}, {and} \bibinfo{person}{Carolyn~Jane Anderson}.} \bibinfo{year}{2023}\natexlab{}.
\newblock \bibinfo{title}{StudentEval: A Benchmark of Student-Written Prompts for Large Language Models of Code}.
\newblock
\newblock
\showeprint[arxiv]{2306.04556}~[cs.LG]


\bibitem[Bai et~al\mbox{.}(2022)]%
        {bai2022constitutional}
\bibfield{author}{\bibinfo{person}{Yuntao Bai}, \bibinfo{person}{Saurav Kadavath}, \bibinfo{person}{Sandipan Kundu}, \bibinfo{person}{Amanda Askell}, \bibinfo{person}{Jackson Kernion}, \bibinfo{person}{Andy Jones}, \bibinfo{person}{Anna Chen}, \bibinfo{person}{Anna Goldie}, \bibinfo{person}{Azalia Mirhoseini}, \bibinfo{person}{Cameron McKinnon}, {et~al\mbox{.}}} \bibinfo{year}{2022}\natexlab{}.
\newblock \showarticletitle{Constitutional ai: Harmlessness from ai feedback}.
\newblock \bibinfo{journal}{\emph{arXiv preprint arXiv:2212.08073}} (\bibinfo{year}{2022}).
\newblock


\bibitem[Bhatt et~al\mbox{.}(2023)]%
        {bhatt2023purple}
\bibfield{author}{\bibinfo{person}{Manish Bhatt}, \bibinfo{person}{Sahana Chennabasappa}, \bibinfo{person}{Cyrus Nikolaidis}, \bibinfo{person}{Shengye Wan}, \bibinfo{person}{Ivan Evtimov}, \bibinfo{person}{Dominik Gabi}, \bibinfo{person}{Daniel Song}, \bibinfo{person}{Faizan Ahmad}, \bibinfo{person}{Cornelius Aschermann}, \bibinfo{person}{Lorenzo Fontana}, {et~al\mbox{.}}} \bibinfo{year}{2023}\natexlab{}.
\newblock \showarticletitle{Purple llama cyberseceval: A secure coding benchmark for language models}.
\newblock \bibinfo{journal}{\emph{arXiv preprint arXiv:2312.04724}} (\bibinfo{year}{2023}).
\newblock


\bibitem[Cassano et~al\mbox{.}(2023)]%
        {cassano2023multipl}
\bibfield{author}{\bibinfo{person}{Federico Cassano}, \bibinfo{person}{John Gouwar}, \bibinfo{person}{Daniel Nguyen}, \bibinfo{person}{Sydney Nguyen}, \bibinfo{person}{Luna Phipps-Costin}, \bibinfo{person}{Donald Pinckney}, \bibinfo{person}{Ming-Ho Yee}, \bibinfo{person}{Yangtian Zi}, \bibinfo{person}{Carolyn~Jane Anderson}, \bibinfo{person}{Molly~Q Feldman}, {et~al\mbox{.}}} \bibinfo{year}{2023}\natexlab{}.
\newblock \showarticletitle{MultiPL-E: a scalable and polyglot approach to benchmarking neural code generation}.
\newblock \bibinfo{journal}{\emph{IEEE Transactions on Software Engineering}} (\bibinfo{year}{2023}).
\newblock


\bibitem[Chen and others.(2021)]%
        {chen2021evaluating}
\bibfield{author}{\bibinfo{person}{Mark Chen} {and} \bibinfo{person}{others.}} \bibinfo{year}{2021}\natexlab{}.
\newblock \bibinfo{title}{Evaluating Large Language Models Trained on Code}.
\newblock
\newblock
\showeprint[arxiv]{2107.03374}~[cs.LG]


\bibitem[Cui et~al\mbox{.}(2023)]%
        {cui2023ultrafeedback}
\bibfield{author}{\bibinfo{person}{Ganqu Cui}, \bibinfo{person}{Lifan Yuan}, \bibinfo{person}{Ning Ding}, \bibinfo{person}{Guanming Yao}, \bibinfo{person}{Wei Zhu}, \bibinfo{person}{Yuan Ni}, \bibinfo{person}{Guotong Xie}, \bibinfo{person}{Zhiyuan Liu}, {and} \bibinfo{person}{Maosong Sun}.} \bibinfo{year}{2023}\natexlab{}.
\newblock \showarticletitle{Ultrafeedback: Boosting language models with high-quality feedback}.
\newblock \bibinfo{journal}{\emph{arXiv preprint arXiv:2310.01377}} (\bibinfo{year}{2023}).
\newblock


\bibitem[Dettmers et~al\mbox{.}(2023)]%
        {dettmers2023qlora}
\bibfield{author}{\bibinfo{person}{Tim Dettmers}, \bibinfo{person}{Artidoro Pagnoni}, \bibinfo{person}{Ari Holtzman}, {and} \bibinfo{person}{Luke Zettlemoyer}.} \bibinfo{year}{2023}\natexlab{}.
\newblock \showarticletitle{Qlora: Efficient finetuning of quantized llms}.
\newblock \bibinfo{journal}{\emph{arXiv preprint arXiv:2305.14314}} (\bibinfo{year}{2023}).
\newblock


\bibitem[Ding et~al\mbox{.}(2023)]%
        {ding2023crosscodeeval}
\bibfield{author}{\bibinfo{person}{Yangruibo Ding}, \bibinfo{person}{Zijian Wang}, \bibinfo{person}{Wasi~Uddin Ahmad}, \bibinfo{person}{Hantian Ding}, \bibinfo{person}{Ming Tan}, \bibinfo{person}{Nihal Jain}, \bibinfo{person}{Murali~Krishna Ramanathan}, \bibinfo{person}{Ramesh Nallapati}, \bibinfo{person}{Parminder Bhatia}, \bibinfo{person}{Dan Roth}, {et~al\mbox{.}}} \bibinfo{year}{2023}\natexlab{}.
\newblock \showarticletitle{Crosscodeeval: A diverse and multilingual benchmark for cross-file code completion}.
\newblock \bibinfo{journal}{\emph{arXiv preprint arXiv:2310.11248}} (\bibinfo{year}{2023}).
\newblock


\bibitem[Du et~al\mbox{.}(2023)]%
        {du2023classeval}
\bibfield{author}{\bibinfo{person}{Xueying Du}, \bibinfo{person}{Mingwei Liu}, \bibinfo{person}{Kaixin Wang}, \bibinfo{person}{Hanlin Wang}, \bibinfo{person}{Junwei Liu}, \bibinfo{person}{Yixuan Chen}, \bibinfo{person}{Jiayi Feng}, \bibinfo{person}{Chaofeng Sha}, \bibinfo{person}{Xin Peng}, {and} \bibinfo{person}{Yiling Lou}.} \bibinfo{year}{2023}\natexlab{}.
\newblock \showarticletitle{Classeval: A manually-crafted benchmark for evaluating llms on class-level code generation}.
\newblock \bibinfo{journal}{\emph{arXiv preprint arXiv:2308.01861}} (\bibinfo{year}{2023}).
\newblock


\bibitem[Guo et~al\mbox{.}(2024)]%
        {guo2024deepseek}
\bibfield{author}{\bibinfo{person}{Daya Guo}, \bibinfo{person}{Qihao Zhu}, \bibinfo{person}{Dejian Yang}, \bibinfo{person}{Zhenda Xie}, \bibinfo{person}{Kai Dong}, \bibinfo{person}{Wentao Zhang}, \bibinfo{person}{Guanting Chen}, \bibinfo{person}{Xiao Bi}, \bibinfo{person}{Y Wu}, \bibinfo{person}{YK Li}, {et~al\mbox{.}}} \bibinfo{year}{2024}\natexlab{}.
\newblock \showarticletitle{DeepSeek-Coder: When the Large Language Model Meets Programming--The Rise of Code Intelligence}.
\newblock \bibinfo{journal}{\emph{arXiv preprint arXiv:2401.14196}} (\bibinfo{year}{2024}).
\newblock


\bibitem[Hendrycks et~al\mbox{.}(2021)]%
        {hendrycksapps2021}
\bibfield{author}{\bibinfo{person}{Dan Hendrycks}, \bibinfo{person}{Steven Basart}, \bibinfo{person}{Saurav Kadavath}, \bibinfo{person}{Mantas Mazeika}, \bibinfo{person}{Akul Arora}, \bibinfo{person}{Ethan Guo}, \bibinfo{person}{Collin Burns}, \bibinfo{person}{Samir Puranik}, \bibinfo{person}{Horace He}, \bibinfo{person}{Dawn Song}, {and} \bibinfo{person}{Jacob Steinhardt}.} \bibinfo{year}{2021}\natexlab{}.
\newblock \showarticletitle{Measuring Coding Challenge Competence With APPS}.
\newblock \bibinfo{journal}{\emph{NeurIPS}} (\bibinfo{year}{2021}).
\newblock


\bibitem[Huang et~al\mbox{.}(2024)]%
        {huang2024effibench}
\bibfield{author}{\bibinfo{person}{Dong Huang}, \bibinfo{person}{Jie~M Zhang}, \bibinfo{person}{Yuhao Qing}, {and} \bibinfo{person}{Heming Cui}.} \bibinfo{year}{2024}\natexlab{}.
\newblock \showarticletitle{EffiBench: Benchmarking the Efficiency of Automatically Generated Code}.
\newblock \bibinfo{journal}{\emph{arXiv preprint arXiv:2402.02037}} (\bibinfo{year}{2024}).
\newblock


\bibitem[Husain et~al\mbox{.}(2019)]%
        {husain2019codesearchnet}
\bibfield{author}{\bibinfo{person}{Hamel Husain}, \bibinfo{person}{Ho-Hsiang Wu}, \bibinfo{person}{Tiferet Gazit}, \bibinfo{person}{Miltiadis Allamanis}, {and} \bibinfo{person}{Marc Brockschmidt}.} \bibinfo{year}{2019}\natexlab{}.
\newblock \showarticletitle{Codesearchnet challenge: Evaluating the state of semantic code search}.
\newblock \bibinfo{journal}{\emph{arXiv preprint arXiv:1909.09436}} (\bibinfo{year}{2019}).
\newblock


\bibitem[Jiang et~al\mbox{.}(2023)]%
        {jiang2023mistral}
\bibfield{author}{\bibinfo{person}{Albert~Q Jiang}, \bibinfo{person}{Alexandre Sablayrolles}, \bibinfo{person}{Arthur Mensch}, \bibinfo{person}{Chris Bamford}, \bibinfo{person}{Devendra~Singh Chaplot}, \bibinfo{person}{Diego de~las Casas}, \bibinfo{person}{Florian Bressand}, \bibinfo{person}{Gianna Lengyel}, \bibinfo{person}{Guillaume Lample}, \bibinfo{person}{Lucile Saulnier}, {et~al\mbox{.}}} \bibinfo{year}{2023}\natexlab{}.
\newblock \showarticletitle{Mistral 7B}.
\newblock \bibinfo{journal}{\emph{arXiv preprint arXiv:2310.06825}} (\bibinfo{year}{2023}).
\newblock


\bibitem[Jiao et~al\mbox{.}(2023)]%
        {jiao2023evaluation}
\bibfield{author}{\bibinfo{person}{Mingsheng Jiao}, \bibinfo{person}{Tingrui Yu}, \bibinfo{person}{Xuan Li}, \bibinfo{person}{Guanjie Qiu}, \bibinfo{person}{Xiaodong Gu}, {and} \bibinfo{person}{Beijun Shen}.} \bibinfo{year}{2023}\natexlab{}.
\newblock \showarticletitle{On the Evaluation of Neural Code Translation: Taxonomy and Benchmark}. In \bibinfo{booktitle}{\emph{2023 38th IEEE/ACM International Conference on Automated Software Engineering (ASE)}}. IEEE, \bibinfo{pages}{1529--1541}.
\newblock


\bibitem[Khan et~al\mbox{.}(2023)]%
        {khan2023xcodeeval}
\bibfield{author}{\bibinfo{person}{Mohammad Abdullah~Matin Khan}, \bibinfo{person}{M~Saiful Bari}, \bibinfo{person}{Xuan~Long Do}, \bibinfo{person}{Weishi Wang}, \bibinfo{person}{Md~Rizwan Parvez}, {and} \bibinfo{person}{Shafiq Joty}.} \bibinfo{year}{2023}\natexlab{}.
\newblock \showarticletitle{xCodeEval: A Large Scale Multilingual Multitask Benchmark for Code Understanding, Generation, Translation and Retrieval}.
\newblock \bibinfo{journal}{\emph{arXiv preprint arXiv:2303.03004}} (\bibinfo{year}{2023}).
\newblock


\bibitem[Kim et~al\mbox{.}(2023)]%
        {kim2023prometheus}
\bibfield{author}{\bibinfo{person}{Seungone Kim}, \bibinfo{person}{Jamin Shin}, \bibinfo{person}{Yejin Cho}, \bibinfo{person}{Joel Jang}, \bibinfo{person}{Shayne Longpre}, \bibinfo{person}{Hwaran Lee}, \bibinfo{person}{Sangdoo Yun}, \bibinfo{person}{Seongjin Shin}, \bibinfo{person}{Sungdong Kim}, \bibinfo{person}{James Thorne}, {et~al\mbox{.}}} \bibinfo{year}{2023}\natexlab{}.
\newblock \showarticletitle{Prometheus: Inducing fine-grained evaluation capability in language models}.
\newblock \bibinfo{journal}{\emph{arXiv preprint arXiv:2310.08491}} (\bibinfo{year}{2023}).
\newblock


\bibitem[Lai et~al\mbox{.}(2023)]%
        {lai2023ds}
\bibfield{author}{\bibinfo{person}{Yuhang Lai}, \bibinfo{person}{Chengxi Li}, \bibinfo{person}{Yiming Wang}, \bibinfo{person}{Tianyi Zhang}, \bibinfo{person}{Ruiqi Zhong}, \bibinfo{person}{Luke Zettlemoyer}, \bibinfo{person}{Wen-tau Yih}, \bibinfo{person}{Daniel Fried}, \bibinfo{person}{Sida Wang}, {and} \bibinfo{person}{Tao Yu}.} \bibinfo{year}{2023}\natexlab{}.
\newblock \showarticletitle{DS-1000: A natural and reliable benchmark for data science code generation}. In \bibinfo{booktitle}{\emph{International Conference on Machine Learning}}. PMLR, \bibinfo{pages}{18319--18345}.
\newblock


\bibitem[Leandro et~al\mbox{.}(2020)]%
        {trl}
\bibfield{author}{\bibinfo{person}{Von~Werra Leandro}, \bibinfo{person}{Belkada Younes}, \bibinfo{person}{Tunstall Lewis}, \bibinfo{person}{Beeching Edward}, \bibinfo{person}{Thrush Tristan}, \bibinfo{person}{Lambert Nathan}, {and} \bibinfo{person}{Huang Shengyi}.} \bibinfo{year}{2020}\natexlab{}.
\newblock \bibinfo{title}{TRL: Transformer reinforcement learning}.
\newblock \bibinfo{howpublished}{\url{https://huggingface.co/docs/trl/en/index}}.
\newblock


\bibitem[Lee et~al\mbox{.}(2023)]%
        {lee2023rlaif}
\bibfield{author}{\bibinfo{person}{Harrison Lee}, \bibinfo{person}{Samrat Phatale}, \bibinfo{person}{Hassan Mansoor}, \bibinfo{person}{Kellie Lu}, \bibinfo{person}{Thomas Mesnard}, \bibinfo{person}{Colton Bishop}, \bibinfo{person}{Victor Carbune}, {and} \bibinfo{person}{Abhinav Rastogi}.} \bibinfo{year}{2023}\natexlab{}.
\newblock \showarticletitle{Rlaif: Scaling reinforcement learning from human feedback with ai feedback}.
\newblock \bibinfo{journal}{\emph{arXiv preprint arXiv:2309.00267}} (\bibinfo{year}{2023}).
\newblock


\bibitem[Li et~al\mbox{.}(2023a)]%
        {alpaca_eval}
\bibfield{author}{\bibinfo{person}{Xuechen Li}, \bibinfo{person}{Tianyi Zhang}, \bibinfo{person}{Yann Dubois}, \bibinfo{person}{Rohan Taori}, \bibinfo{person}{Ishaan Gulrajani}, \bibinfo{person}{Carlos Guestrin}, \bibinfo{person}{Percy Liang}, {and} \bibinfo{person}{Tatsunori~B. Hashimoto}.} \bibinfo{year}{2023}\natexlab{a}.
\newblock \bibinfo{title}{AlpacaEval: An Automatic Evaluator of Instruction-following Models}.
\newblock \bibinfo{howpublished}{\url{https://github.com/tatsu-lab/alpaca_eval}}.
\newblock


\bibitem[Li et~al\mbox{.}(2023b)]%
        {li2023alpacaeval}
\bibfield{author}{\bibinfo{person}{Xuechen Li}, \bibinfo{person}{Tianyi Zhang}, \bibinfo{person}{Yann Dubois}, \bibinfo{person}{Rohan Taori}, \bibinfo{person}{Ishaan Gulrajani}, \bibinfo{person}{Carlos Guestrin}, \bibinfo{person}{Percy Liang}, {and} \bibinfo{person}{Tatsunori~B Hashimoto}.} \bibinfo{year}{2023}\natexlab{b}.
\newblock \showarticletitle{Alpacaeval: An automatic evaluator of instruction-following models}.
\newblock \bibinfo{journal}{\emph{GitHub repository}} (\bibinfo{year}{2023}).
\newblock


\bibitem[Liu et~al\mbox{.}(2023b)]%
        {liu2023your}
\bibfield{author}{\bibinfo{person}{Jiawei Liu}, \bibinfo{person}{Chunqiu~Steven Xia}, \bibinfo{person}{Yuyao Wang}, {and} \bibinfo{person}{Lingming Zhang}.} \bibinfo{year}{2023}\natexlab{b}.
\newblock \showarticletitle{Is your code generated by chatgpt really correct? rigorous evaluation of large language models for code generation}.
\newblock \bibinfo{journal}{\emph{arXiv preprint arXiv:2305.01210}} (\bibinfo{year}{2023}).
\newblock


\bibitem[Liu et~al\mbox{.}(2023a)]%
        {liu2023refining}
\bibfield{author}{\bibinfo{person}{Yue Liu}, \bibinfo{person}{Thanh Le-Cong}, \bibinfo{person}{Ratnadira Widyasari}, \bibinfo{person}{Chakkrit Tantithamthavorn}, \bibinfo{person}{Li Li}, \bibinfo{person}{Xuan-Bach~D Le}, {and} \bibinfo{person}{David Lo}.} \bibinfo{year}{2023}\natexlab{a}.
\newblock \showarticletitle{Refining ChatGPT-generated code: Characterizing and mitigating code quality issues}.
\newblock \bibinfo{journal}{\emph{ACM Transactions on Software Engineering and Methodology}} (\bibinfo{year}{2023}).
\newblock


\bibitem[Lu et~al\mbox{.}(2021)]%
        {lu2021codexglue}
\bibfield{author}{\bibinfo{person}{Shuai Lu}, \bibinfo{person}{Daya Guo}, \bibinfo{person}{Shuo Ren}, \bibinfo{person}{Junjie Huang}, \bibinfo{person}{Alexey Svyatkovskiy}, \bibinfo{person}{Ambrosio Blanco}, \bibinfo{person}{Colin Clement}, \bibinfo{person}{Dawn Drain}, \bibinfo{person}{Daxin Jiang}, \bibinfo{person}{Duyu Tang}, {et~al\mbox{.}}} \bibinfo{year}{2021}\natexlab{}.
\newblock \showarticletitle{Codexglue: A machine learning benchmark dataset for code understanding and generation}.
\newblock \bibinfo{journal}{\emph{arXiv preprint arXiv:2102.04664}} (\bibinfo{year}{2021}).
\newblock


\bibitem[Luo et~al\mbox{.}(2023)]%
        {luo2023wizardcoder}
\bibfield{author}{\bibinfo{person}{Ziyang Luo}, \bibinfo{person}{Can Xu}, \bibinfo{person}{Pu Zhao}, \bibinfo{person}{Qingfeng Sun}, \bibinfo{person}{Xiubo Geng}, \bibinfo{person}{Wenxiang Hu}, \bibinfo{person}{Chongyang Tao}, \bibinfo{person}{Jing Ma}, \bibinfo{person}{Qingwei Lin}, {and} \bibinfo{person}{Daxin Jiang}.} \bibinfo{year}{2023}\natexlab{}.
\newblock \showarticletitle{WizardCoder: Empowering Code Large Language Models with Evol-Instruct}.
\newblock \bibinfo{journal}{\emph{arXiv preprint arXiv:2306.08568}} (\bibinfo{year}{2023}).
\newblock


\bibitem[Muennighoff et~al\mbox{.}(2023)]%
        {muennighoff2023octopack}
\bibfield{author}{\bibinfo{person}{Niklas Muennighoff}, \bibinfo{person}{Qian Liu}, \bibinfo{person}{Armel Zebaze}, \bibinfo{person}{Qinkai Zheng}, \bibinfo{person}{Binyuan Hui}, \bibinfo{person}{Terry~Yue Zhuo}, \bibinfo{person}{Swayam Singh}, \bibinfo{person}{Xiangru Tang}, \bibinfo{person}{Leandro Von~Werra}, {and} \bibinfo{person}{Shayne Longpre}.} \bibinfo{year}{2023}\natexlab{}.
\newblock \showarticletitle{Octopack: Instruction tuning code large language models}.
\newblock \bibinfo{journal}{\emph{arXiv preprint arXiv:2308.07124}} (\bibinfo{year}{2023}).
\newblock


\bibitem[Mukherjee et~al\mbox{.}(2023)]%
        {mukherjee2023orca}
\bibfield{author}{\bibinfo{person}{Subhabrata Mukherjee}, \bibinfo{person}{Arindam Mitra}, \bibinfo{person}{Ganesh Jawahar}, \bibinfo{person}{Sahaj Agarwal}, \bibinfo{person}{Hamid Palangi}, {and} \bibinfo{person}{Ahmed Awadallah}.} \bibinfo{year}{2023}\natexlab{}.
\newblock \showarticletitle{Orca: Progressive learning from complex explanation traces of gpt-4}.
\newblock \bibinfo{journal}{\emph{arXiv preprint arXiv:2306.02707}} (\bibinfo{year}{2023}).
\newblock


\bibitem[Niu et~al\mbox{.}(2023)]%
        {niu2023crosscodebench}
\bibfield{author}{\bibinfo{person}{Changan Niu}, \bibinfo{person}{Chuanyi Li}, \bibinfo{person}{Vincent Ng}, {and} \bibinfo{person}{Bin Luo}.} \bibinfo{year}{2023}\natexlab{}.
\newblock \showarticletitle{CrossCodeBench: Benchmarking Cross-Task Generalization of Source Code Models}.
\newblock \bibinfo{journal}{\emph{arXiv preprint arXiv:2302.04030}} (\bibinfo{year}{2023}).
\newblock


\bibitem[Ouyang et~al\mbox{.}(2022)]%
        {ouyang2022training}
\bibfield{author}{\bibinfo{person}{Long Ouyang}, \bibinfo{person}{Jeff Wu}, \bibinfo{person}{Xu Jiang}, \bibinfo{person}{Diogo Almeida}, \bibinfo{person}{Carroll~L Wainwright}, \bibinfo{person}{Pamela Mishkin}, \bibinfo{person}{Chong Zhang}, \bibinfo{person}{Sandhini Agarwal}, \bibinfo{person}{Katarina Slama}, \bibinfo{person}{Alex Ray}, {et~al\mbox{.}}} \bibinfo{year}{2022}\natexlab{}.
\newblock \showarticletitle{Training language models to follow instructions with human feedback, 2022}.
\newblock \bibinfo{journal}{\emph{URL https://arxiv. org/abs/2203.02155}}  \bibinfo{volume}{13} (\bibinfo{year}{2022}).
\newblock


\bibitem[Pan et~al\mbox{.}(2023)]%
        {pan2023understanding}
\bibfield{author}{\bibinfo{person}{Rangeet Pan}, \bibinfo{person}{Ali~Reza Ibrahimzada}, \bibinfo{person}{Rahul Krishna}, \bibinfo{person}{Divya Sankar}, \bibinfo{person}{Lambert~Pouguem Wassi}, \bibinfo{person}{Michele Merler}, \bibinfo{person}{Boris Sobolev}, \bibinfo{person}{Raju Pavuluri}, \bibinfo{person}{Saurabh Sinha}, {and} \bibinfo{person}{Reyhaneh Jabbarvand}.} \bibinfo{year}{2023}\natexlab{}.
\newblock \showarticletitle{Understanding the effectiveness of large language models in code translation}.
\newblock \bibinfo{journal}{\emph{arXiv preprint arXiv:2308.03109}} (\bibinfo{year}{2023}).
\newblock


\bibitem[Puri et~al\mbox{.}(2021)]%
        {puri2021codenet}
\bibfield{author}{\bibinfo{person}{Ruchir Puri}, \bibinfo{person}{David~S Kung}, \bibinfo{person}{Geert Janssen}, \bibinfo{person}{Wei Zhang}, \bibinfo{person}{Giacomo Domeniconi}, \bibinfo{person}{Vladimir Zolotov}, \bibinfo{person}{Julian Dolby}, \bibinfo{person}{Jie Chen}, \bibinfo{person}{Mihir Choudhury}, \bibinfo{person}{Lindsey Decker}, {et~al\mbox{.}}} \bibinfo{year}{2021}\natexlab{}.
\newblock \showarticletitle{Codenet: A large-scale ai for code dataset for learning a diversity of coding tasks}.
\newblock \bibinfo{journal}{\emph{arXiv preprint arXiv:2105.12655}} (\bibinfo{year}{2021}).
\newblock


\bibitem[Rafailov et~al\mbox{.}(2023)]%
        {rafailov2023direct}
\bibfield{author}{\bibinfo{person}{Rafael Rafailov}, \bibinfo{person}{Archit Sharma}, \bibinfo{person}{Eric Mitchell}, \bibinfo{person}{Stefano Ermon}, \bibinfo{person}{Christopher~D Manning}, {and} \bibinfo{person}{Chelsea Finn}.} \bibinfo{year}{2023}\natexlab{}.
\newblock \showarticletitle{Direct preference optimization: Your language model is secretly a reward model}.
\newblock \bibinfo{journal}{\emph{arXiv preprint arXiv:2305.18290}} (\bibinfo{year}{2023}).
\newblock


\bibitem[Raychev et~al\mbox{.}(2016a)]%
        {raychev2016probabilistic}
\bibfield{author}{\bibinfo{person}{Veselin Raychev}, \bibinfo{person}{Pavol Bielik}, {and} \bibinfo{person}{Martin Vechev}.} \bibinfo{year}{2016}\natexlab{a}.
\newblock \showarticletitle{Probabilistic model for code with decision trees}.
\newblock \bibinfo{journal}{\emph{ACM SIGPLAN Notices}} \bibinfo{volume}{51}, \bibinfo{number}{10} (\bibinfo{year}{2016}), \bibinfo{pages}{731--747}.
\newblock


\bibitem[Raychev et~al\mbox{.}(2016b)]%
        {raychev2016learning}
\bibfield{author}{\bibinfo{person}{Veselin Raychev}, \bibinfo{person}{Pavol Bielik}, \bibinfo{person}{Martin Vechev}, {and} \bibinfo{person}{Andreas Krause}.} \bibinfo{year}{2016}\natexlab{b}.
\newblock \showarticletitle{Learning programs from noisy data}.
\newblock \bibinfo{journal}{\emph{ACM Sigplan Notices}} \bibinfo{volume}{51}, \bibinfo{number}{1} (\bibinfo{year}{2016}), \bibinfo{pages}{761--774}.
\newblock


\bibitem[Roziere et~al\mbox{.}(2023)]%
        {roziere2023code}
\bibfield{author}{\bibinfo{person}{Baptiste Roziere}, \bibinfo{person}{Jonas Gehring}, \bibinfo{person}{Fabian Gloeckle}, \bibinfo{person}{Sten Sootla}, \bibinfo{person}{Itai Gat}, \bibinfo{person}{Xiaoqing~Ellen Tan}, \bibinfo{person}{Yossi Adi}, \bibinfo{person}{Jingyu Liu}, \bibinfo{person}{Tal Remez}, \bibinfo{person}{J{\'e}r{\'e}my Rapin}, {et~al\mbox{.}}} \bibinfo{year}{2023}\natexlab{}.
\newblock \showarticletitle{Code llama: Open foundation models for code}.
\newblock \bibinfo{journal}{\emph{arXiv preprint arXiv:2308.12950}} (\bibinfo{year}{2023}).
\newblock


\bibitem[Sghaier et~al\mbox{.}(2023)]%
        {sghaier2023unity}
\bibfield{author}{\bibinfo{person}{Oussama~Ben Sghaier}, \bibinfo{person}{Lucas Maes}, {and} \bibinfo{person}{Houari Sahraoui}.} \bibinfo{year}{2023}\natexlab{}.
\newblock \showarticletitle{Unity is Strength: Cross-Task Knowledge Distillation to Improve Code Review Generation}.
\newblock \bibinfo{journal}{\emph{arXiv preprint arXiv:2309.03362}} (\bibinfo{year}{2023}).
\newblock


\bibitem[Sghaier and Sahraoui(2023)]%
        {sghaier2023multi}
\bibfield{author}{\bibinfo{person}{Oussama~Ben Sghaier} {and} \bibinfo{person}{Houari Sahraoui}.} \bibinfo{year}{2023}\natexlab{}.
\newblock \showarticletitle{A Multi-Step Learning Approach to Assist Code Review}. In \bibinfo{booktitle}{\emph{2023 IEEE International Conference on Software Analysis, Evolution and Reengineering (SANER)}}. IEEE, \bibinfo{pages}{450--460}.
\newblock


\bibitem[Sghaier and Sahraoui(2024)]%
        {sghaier2024improving}
\bibfield{author}{\bibinfo{person}{Oussama~Ben Sghaier} {and} \bibinfo{person}{Houari Sahraoui}.} \bibinfo{year}{2024}\natexlab{}.
\newblock \showarticletitle{Improving the Learning of Code Review Successive Tasks with Cross-Task Knowledge Distillation}.
\newblock \bibinfo{journal}{\emph{arXiv preprint arXiv:2402.02063}} (\bibinfo{year}{2024}).
\newblock


\bibitem[Siddiq et~al\mbox{.}(2023)]%
        {siddiq2023lightweight}
\bibfield{author}{\bibinfo{person}{Mohammed~Latif Siddiq}, \bibinfo{person}{Beatrice Casey}, {and} \bibinfo{person}{Joanna Santos}.} \bibinfo{year}{2023}\natexlab{}.
\newblock \showarticletitle{A Lightweight Framework for High-Quality Code Generation}.
\newblock \bibinfo{journal}{\emph{arXiv preprint arXiv:2307.08220}} (\bibinfo{year}{2023}).
\newblock


\bibitem[Silva et~al\mbox{.}(2023)]%
        {silva2023repairllama}
\bibfield{author}{\bibinfo{person}{Andr{\'e} Silva}, \bibinfo{person}{Sen Fang}, {and} \bibinfo{person}{Martin Monperrus}.} \bibinfo{year}{2023}\natexlab{}.
\newblock \showarticletitle{RepairLLaMA: Efficient Representations and Fine-Tuned Adapters for Program Repair}.
\newblock \bibinfo{journal}{\emph{arXiv preprint arXiv:2312.15698}} (\bibinfo{year}{2023}).
\newblock


\bibitem[Singhal et~al\mbox{.}(2024)]%
        {singhal2024nofuneval}
\bibfield{author}{\bibinfo{person}{Manav Singhal}, \bibinfo{person}{Tushar Aggarwal}, \bibinfo{person}{Abhijeet Awasthi}, \bibinfo{person}{Nagarajan Natarajan}, {and} \bibinfo{person}{Aditya Kanade}.} \bibinfo{year}{2024}\natexlab{}.
\newblock \showarticletitle{NoFunEval: Funny How Code LMs Falter on Requirements Beyond Functional Correctness}.
\newblock \bibinfo{journal}{\emph{arXiv preprint arXiv:2401.15963}} (\bibinfo{year}{2024}).
\newblock


\bibitem[Sun et~al\mbox{.}(2023)]%
        {sun2023principle}
\bibfield{author}{\bibinfo{person}{Zhiqing Sun}, \bibinfo{person}{Yikang Shen}, \bibinfo{person}{Qinhong Zhou}, \bibinfo{person}{Hongxin Zhang}, \bibinfo{person}{Zhenfang Chen}, \bibinfo{person}{David Cox}, \bibinfo{person}{Yiming Yang}, {and} \bibinfo{person}{Chuang Gan}.} \bibinfo{year}{2023}\natexlab{}.
\newblock \showarticletitle{Principle-driven self-alignment of language models from scratch with minimal human supervision}.
\newblock \bibinfo{journal}{\emph{arXiv preprint arXiv:2305.03047}} (\bibinfo{year}{2023}).
\newblock


\bibitem[Taori et~al\mbox{.}(2023)]%
        {taori2023alpaca}
\bibfield{author}{\bibinfo{person}{Rohan Taori}, \bibinfo{person}{Ishaan Gulrajani}, \bibinfo{person}{Tianyi Zhang}, \bibinfo{person}{Yann Dubois}, \bibinfo{person}{Xuechen Li}, \bibinfo{person}{Carlos Guestrin}, \bibinfo{person}{Percy Liang}, {and} \bibinfo{person}{Tatsunori~B Hashimoto}.} \bibinfo{year}{2023}\natexlab{}.
\newblock \showarticletitle{Alpaca: A strong, replicable instruction-following model}.
\newblock \bibinfo{journal}{\emph{Stanford Center for Research on Foundation Models. https://crfm. stanford. edu/2023/03/13/alpaca. html}} \bibinfo{volume}{3}, \bibinfo{number}{6} (\bibinfo{year}{2023}), \bibinfo{pages}{7}.
\newblock


\bibitem[Touvron et~al\mbox{.}(2023)]%
        {touvron2023llama}
\bibfield{author}{\bibinfo{person}{Hugo Touvron}, \bibinfo{person}{Louis Martin}, \bibinfo{person}{Kevin Stone}, \bibinfo{person}{Peter Albert}, \bibinfo{person}{Amjad Almahairi}, \bibinfo{person}{Yasmine Babaei}, \bibinfo{person}{Nikolay Bashlykov}, \bibinfo{person}{Soumya Batra}, \bibinfo{person}{Prajjwal Bhargava}, \bibinfo{person}{Shruti Bhosale}, {et~al\mbox{.}}} \bibinfo{year}{2023}\natexlab{}.
\newblock \showarticletitle{Llama 2: Open foundation and fine-tuned chat models}.
\newblock \bibinfo{journal}{\emph{arXiv preprint arXiv:2307.09288}} (\bibinfo{year}{2023}).
\newblock


\bibitem[Tunstall et~al\mbox{.}(2023)]%
        {tunstall2023zephyr}
\bibfield{author}{\bibinfo{person}{Lewis Tunstall}, \bibinfo{person}{Edward Beeching}, \bibinfo{person}{Nathan Lambert}, \bibinfo{person}{Nazneen Rajani}, \bibinfo{person}{Kashif Rasul}, \bibinfo{person}{Younes Belkada}, \bibinfo{person}{Shengyi Huang}, \bibinfo{person}{Leandro von Werra}, \bibinfo{person}{Cl{\'e}mentine Fourrier}, \bibinfo{person}{Nathan Habib}, {et~al\mbox{.}}} \bibinfo{year}{2023}\natexlab{}.
\newblock \showarticletitle{Zephyr: Direct distillation of lm alignment}.
\newblock \bibinfo{journal}{\emph{arXiv preprint arXiv:2310.16944}} (\bibinfo{year}{2023}).
\newblock


\bibitem[Wang et~al\mbox{.}(2022b)]%
        {wang2022recode}
\bibfield{author}{\bibinfo{person}{Shiqi Wang}, \bibinfo{person}{Zheng Li}, \bibinfo{person}{Haifeng Qian}, \bibinfo{person}{Chenghao Yang}, \bibinfo{person}{Zijian Wang}, \bibinfo{person}{Mingyue Shang}, \bibinfo{person}{Varun Kumar}, \bibinfo{person}{Samson Tan}, \bibinfo{person}{Baishakhi Ray}, \bibinfo{person}{Parminder Bhatia}, {et~al\mbox{.}}} \bibinfo{year}{2022}\natexlab{b}.
\newblock \showarticletitle{ReCode: Robustness Evaluation of Code Generation Models}.
\newblock \bibinfo{journal}{\emph{arXiv preprint arXiv:2212.10264}} (\bibinfo{year}{2022}).
\newblock


\bibitem[Wang et~al\mbox{.}(2023)]%
        {wang2023shepherd}
\bibfield{author}{\bibinfo{person}{Tianlu Wang}, \bibinfo{person}{Ping Yu}, \bibinfo{person}{Xiaoqing~Ellen Tan}, \bibinfo{person}{Sean O'Brien}, \bibinfo{person}{Ramakanth Pasunuru}, \bibinfo{person}{Jane Dwivedi-Yu}, \bibinfo{person}{Olga Golovneva}, \bibinfo{person}{Luke Zettlemoyer}, \bibinfo{person}{Maryam Fazel-Zarandi}, {and} \bibinfo{person}{Asli Celikyilmaz}.} \bibinfo{year}{2023}\natexlab{}.
\newblock \showarticletitle{Shepherd: A critic for language model generation}.
\newblock \bibinfo{journal}{\emph{arXiv preprint arXiv:2308.04592}} (\bibinfo{year}{2023}).
\newblock


\bibitem[Wang et~al\mbox{.}(2022a)]%
        {wang2022self}
\bibfield{author}{\bibinfo{person}{Yizhong Wang}, \bibinfo{person}{Yeganeh Kordi}, \bibinfo{person}{Swaroop Mishra}, \bibinfo{person}{Alisa Liu}, \bibinfo{person}{Noah~A Smith}, \bibinfo{person}{Daniel Khashabi}, {and} \bibinfo{person}{Hannaneh Hajishirzi}.} \bibinfo{year}{2022}\natexlab{a}.
\newblock \showarticletitle{Self-instruct: Aligning language model with self generated instructions}.
\newblock \bibinfo{journal}{\emph{arXiv preprint arXiv:2212.10560}} (\bibinfo{year}{2022}).
\newblock


\bibitem[Wei et~al\mbox{.}(2022)]%
        {wei2022chain}
\bibfield{author}{\bibinfo{person}{Jason Wei}, \bibinfo{person}{Xuezhi Wang}, \bibinfo{person}{Dale Schuurmans}, \bibinfo{person}{Maarten Bosma}, \bibinfo{person}{Fei Xia}, \bibinfo{person}{Ed Chi}, \bibinfo{person}{Quoc~V Le}, \bibinfo{person}{Denny Zhou}, {et~al\mbox{.}}} \bibinfo{year}{2022}\natexlab{}.
\newblock \showarticletitle{Chain-of-thought prompting elicits reasoning in large language models}.
\newblock \bibinfo{journal}{\emph{Advances in neural information processing systems}}  \bibinfo{volume}{35} (\bibinfo{year}{2022}), \bibinfo{pages}{24824--24837}.
\newblock


\bibitem[Wei et~al\mbox{.}(2023)]%
        {wei2023magicoder}
\bibfield{author}{\bibinfo{person}{Yuxiang Wei}, \bibinfo{person}{Zhe Wang}, \bibinfo{person}{Jiawei Liu}, \bibinfo{person}{Yifeng Ding}, {and} \bibinfo{person}{Lingming Zhang}.} \bibinfo{year}{2023}\natexlab{}.
\newblock \showarticletitle{Magicoder: Source code is all you need}.
\newblock \bibinfo{journal}{\emph{arXiv preprint arXiv:2312.02120}} (\bibinfo{year}{2023}).
\newblock


\bibitem[Weyssow et~al\mbox{.}(2023)]%
        {weyssow2023exploring}
\bibfield{author}{\bibinfo{person}{Martin Weyssow}, \bibinfo{person}{Xin Zhou}, \bibinfo{person}{Kisub Kim}, \bibinfo{person}{David Lo}, {and} \bibinfo{person}{Houari Sahraoui}.} \bibinfo{year}{2023}\natexlab{}.
\newblock \showarticletitle{Exploring parameter-efficient fine-tuning techniques for code generation with large language models}.
\newblock \bibinfo{journal}{\emph{arXiv preprint arXiv:2308.10462}} (\bibinfo{year}{2023}).
\newblock


\bibitem[Xu et~al\mbox{.}(2023)]%
        {xu2023wizardlm}
\bibfield{author}{\bibinfo{person}{Can Xu}, \bibinfo{person}{Qingfeng Sun}, \bibinfo{person}{Kai Zheng}, \bibinfo{person}{Xiubo Geng}, \bibinfo{person}{Pu Zhao}, \bibinfo{person}{Jiazhan Feng}, \bibinfo{person}{Chongyang Tao}, {and} \bibinfo{person}{Daxin Jiang}.} \bibinfo{year}{2023}\natexlab{}.
\newblock \showarticletitle{Wizardlm: Empowering large language models to follow complex instructions}.
\newblock \bibinfo{journal}{\emph{arXiv preprint arXiv:2304.12244}} (\bibinfo{year}{2023}).
\newblock


\bibitem[Xu et~al\mbox{.}(2024)]%
        {xu2024survey}
\bibfield{author}{\bibinfo{person}{Xiaohan Xu}, \bibinfo{person}{Ming Li}, \bibinfo{person}{Chongyang Tao}, \bibinfo{person}{Tao Shen}, \bibinfo{person}{Reynold Cheng}, \bibinfo{person}{Jinyang Li}, \bibinfo{person}{Can Xu}, \bibinfo{person}{Dacheng Tao}, {and} \bibinfo{person}{Tianyi Zhou}.} \bibinfo{year}{2024}\natexlab{}.
\newblock \bibinfo{title}{A Survey on Knowledge Distillation of Large Language Models}.
\newblock
\newblock
\showeprint[arxiv]{2402.13116}~[cs.CL]


\bibitem[Yang et~al\mbox{.}(2024)]%
        {yang2024robustness}
\bibfield{author}{\bibinfo{person}{Zhou Yang}, \bibinfo{person}{Zhensu Sun}, \bibinfo{person}{Terry~Zhuo Yue}, \bibinfo{person}{Premkumar Devanbu}, {and} \bibinfo{person}{David Lo}.} \bibinfo{year}{2024}\natexlab{}.
\newblock \bibinfo{title}{Robustness, Security, Privacy, Explainability, Efficiency, and Usability of Large Language Models for Code}.
\newblock
\newblock
\showeprint[arxiv]{2403.07506}~[cs.SE]


\bibitem[Ye et~al\mbox{.}(2021)]%
        {ye2021comprehensive}
\bibfield{author}{\bibinfo{person}{He Ye}, \bibinfo{person}{Matias Martinez}, \bibinfo{person}{Thomas Durieux}, {and} \bibinfo{person}{Martin Monperrus}.} \bibinfo{year}{2021}\natexlab{}.
\newblock \showarticletitle{A comprehensive study of automatic program repair on the QuixBugs benchmark}.
\newblock \bibinfo{journal}{\emph{Journal of Systems and Software}}  \bibinfo{volume}{171} (\bibinfo{year}{2021}), \bibinfo{pages}{110825}.
\newblock


\bibitem[Yeti{\c{s}}tiren et~al\mbox{.}(2023)]%
        {yeticstiren2023evaluating}
\bibfield{author}{\bibinfo{person}{Burak Yeti{\c{s}}tiren}, \bibinfo{person}{I{\c{s}}{\i}k {\"O}zsoy}, \bibinfo{person}{Miray Ayerdem}, {and} \bibinfo{person}{Eray T{\"u}z{\"u}n}.} \bibinfo{year}{2023}\natexlab{}.
\newblock \showarticletitle{Evaluating the Code Quality of AI-Assisted Code Generation Tools: An Empirical Study on GitHub Copilot, Amazon CodeWhisperer, and ChatGPT}.
\newblock \bibinfo{journal}{\emph{arXiv preprint arXiv:2304.10778}} (\bibinfo{year}{2023}).
\newblock


\bibitem[Yu et~al\mbox{.}(2024)]%
        {yu2024codereval}
\bibfield{author}{\bibinfo{person}{Hao Yu}, \bibinfo{person}{Bo Shen}, \bibinfo{person}{Dezhi Ran}, \bibinfo{person}{Jiaxin Zhang}, \bibinfo{person}{Qi Zhang}, \bibinfo{person}{Yuchi Ma}, \bibinfo{person}{Guangtai Liang}, \bibinfo{person}{Ying Li}, \bibinfo{person}{Qianxiang Wang}, {and} \bibinfo{person}{Tao Xie}.} \bibinfo{year}{2024}\natexlab{}.
\newblock \showarticletitle{Codereval: A benchmark of pragmatic code generation with generative pre-trained models}. In \bibinfo{booktitle}{\emph{Proceedings of the 46th IEEE/ACM International Conference on Software Engineering}}. \bibinfo{pages}{1--12}.
\newblock


\bibitem[Zheng et~al\mbox{.}(2023a)]%
        {zheng2023judging}
\bibfield{author}{\bibinfo{person}{Lianmin Zheng}, \bibinfo{person}{Wei-Lin Chiang}, \bibinfo{person}{Ying Sheng}, \bibinfo{person}{Siyuan Zhuang}, \bibinfo{person}{Zhanghao Wu}, \bibinfo{person}{Yonghao Zhuang}, \bibinfo{person}{Zi Lin}, \bibinfo{person}{Zhuohan Li}, \bibinfo{person}{Dacheng Li}, \bibinfo{person}{Eric Xing}, {et~al\mbox{.}}} \bibinfo{year}{2023}\natexlab{a}.
\newblock \showarticletitle{Judging LLM-as-a-judge with MT-Bench and Chatbot Arena}.
\newblock \bibinfo{journal}{\emph{arXiv preprint arXiv:2306.05685}} (\bibinfo{year}{2023}).
\newblock


\bibitem[Zheng et~al\mbox{.}(2023b)]%
        {zheng2023codegeex}
\bibfield{author}{\bibinfo{person}{Qinkai Zheng}, \bibinfo{person}{Xiao Xia}, \bibinfo{person}{Xu Zou}, \bibinfo{person}{Yuxiao Dong}, \bibinfo{person}{Shan Wang}, \bibinfo{person}{Yufei Xue}, \bibinfo{person}{Zihan Wang}, \bibinfo{person}{Lei Shen}, \bibinfo{person}{Andi Wang}, \bibinfo{person}{Yang Li}, {et~al\mbox{.}}} \bibinfo{year}{2023}\natexlab{b}.
\newblock \showarticletitle{Codegeex: A pre-trained model for code generation with multilingual evaluations on humaneval-x}.
\newblock \bibinfo{journal}{\emph{arXiv preprint arXiv:2303.17568}} (\bibinfo{year}{2023}).
\newblock


\bibitem[Zhou et~al\mbox{.}(2022)]%
        {zhou2022docprompting}
\bibfield{author}{\bibinfo{person}{Shuyan Zhou}, \bibinfo{person}{Uri Alon}, \bibinfo{person}{Frank~F Xu}, \bibinfo{person}{Zhengbao Jiang}, {and} \bibinfo{person}{Graham Neubig}.} \bibinfo{year}{2022}\natexlab{}.
\newblock \showarticletitle{Docprompting: Generating code by retrieving the docs}. In \bibinfo{booktitle}{\emph{The Eleventh International Conference on Learning Representations}}.
\newblock


\bibitem[Zhou et~al\mbox{.}(2023)]%
        {zhou2023generation}
\bibfield{author}{\bibinfo{person}{Xin Zhou}, \bibinfo{person}{Kisub Kim}, \bibinfo{person}{Bowen Xu}, \bibinfo{person}{DongGyun Han}, \bibinfo{person}{Junda He}, {and} \bibinfo{person}{David Lo}.} \bibinfo{year}{2023}\natexlab{}.
\newblock \showarticletitle{Generation-based Code Review Automation: How Far Are We?}
\newblock \bibinfo{journal}{\emph{arXiv preprint arXiv:2303.07221}} (\bibinfo{year}{2023}).
\newblock


\bibitem[Zhu et~al\mbox{.}(2022)]%
        {zhu2022xlcost}
\bibfield{author}{\bibinfo{person}{Ming Zhu}, \bibinfo{person}{Aneesh Jain}, \bibinfo{person}{Karthik Suresh}, \bibinfo{person}{Roshan Ravindran}, \bibinfo{person}{Sindhu Tipirneni}, {and} \bibinfo{person}{Chandan~K Reddy}.} \bibinfo{year}{2022}\natexlab{}.
\newblock \showarticletitle{Xlcost: A benchmark dataset for cross-lingual code intelligence}.
\newblock \bibinfo{journal}{\emph{arXiv preprint arXiv:2206.08474}} (\bibinfo{year}{2022}).
\newblock


\bibitem[Zhuo(2024)]%
        {zhuo2024ice}
\bibfield{author}{\bibinfo{person}{Terry~Yue Zhuo}.} \bibinfo{year}{2024}\natexlab{}.
\newblock \showarticletitle{ICE-Score: Instructing Large Language Models to Evaluate Code}. In \bibinfo{booktitle}{\emph{Findings of the Association for Computational Linguistics: EACL 2024}}. \bibinfo{pages}{2232--2242}.
\newblock


\end{thebibliography}

\appendix

\section{Assessment Templates}

\renewcommand{\arraystretch}{1}
\setlength{\arrayrulewidth}{.5pt}
\setlength{\tabcolsep}{3pt}
\begin{table}[!t]
\centering
\footnotesize
\caption{Instruction following assessment template.}
\label{tab:instruction_judging_template}
\vspace{-1em}
    \begin{tabular*}{\linewidth}{p{.99\linewidth}}
    \toprule
Evaluate the assistant's fidelity to provided instructions. Assess how accurately the assistant's responses align with user directives, noting any deviations and their justification.

\textbf{Evaluation Criteria}: \\
\:\: \textit{Precision in Following Instructions}: Does the assistant adhere to the specifics of the provided instructions? \\
\:\: \textit{Justification for Deviations}: If deviations occur, are they justified by critical necessity or explicit user request? \\
\:\: \textit{Alignment with User Directives}: How well do the assistant's responses  match the user’s specified needs and expectations? \\
\:\: \textit{Necessity of Deviations}: Are any deviations from instructions made only in situations deemed absolutely necessary or upon direct user request? \\
\textbf{Scoring}: Rate outputs on a scale of 1 to 5: \\
\:\:1. \textit{Non-Compliant}: The assistant frequently deviates from instructions without necessity or user consent. \\
\:\:2. \textit{Minimally Compliant}: The assistant shows some adherence to instructions but deviates without strong justification. \\
\:\:3. \textit{Moderately Compliant}: The assistant generally follows instructions, with deviations occurring but justified by necessity or user request. \\
\:\:4. \textit{Highly Compliant}: The assistant closely follows instructions, with few deviations, all of which are well justified. \\
\:\:5. \textit{Fully Compliant}: The assistant exhibits perfect adherence to instructions, with deviations only when critically necessary and explicitly approved by the user. \\

    \arrayrulecolor{black}
    \bottomrule
    \end{tabular*}
\end{table}

\renewcommand{\arraystretch}{1}
\setlength{\arrayrulewidth}{.5pt}
\setlength{\tabcolsep}{3pt}
\begin{table}[!t]
\centering
\footnotesize
\caption{Code explanation assessment template.}
\label{tab:explanation_judging_template}
\vspace{-1em}
    \begin{tabular*}{\linewidth}{p{.99\linewidth}}
    \toprule
Evaluate the clarity and depth of explanations accompanying code segments. Assess how well the explanation helps in understanding the code's purpose, logic, and design choices.

\textbf{Evaluation Criteria}: \\
\:\: \textit{Clarity}: How easy is it to understand the explanation? \\
\:\: \textit{Depth}: Does the explanation cover the logic, structure, and decisions behind the code? \\
\:\: \textit{Relevance}: Is the explanation relevant to the code's purpose and design philosophy? \\
\:\: \textit{Accessibility}: Can a broad audience understand the explanation, regardless of their technical background? \\
\textbf{Scoring}: Rate outputs on a scale of 1 to 5: \\
\:\:1. \textit{Inadequate}: Explanation is unclear, superficial, or missing. \\
\:\:2. \textit{Basic}: Explanation covers fundamental points but lacks depth or clarity. \\
\:\:3. \textit{Good}: Explanation is clear and somewhat detailed, but may miss some deeper insights. \\
\:\:4. \textit{Very Good}: Explanation is clear, detailed, and covers most logic, structure, and decisions. \\
\:\:5. \textit{Excellent}: Explanation is exceptionally clear, in-depth, and makes the code's purpose, design, and logic accessible to all users. \\

    \arrayrulecolor{black}
    \bottomrule
    \end{tabular*}
\end{table}

\renewcommand{\arraystretch}{1}
\setlength{\arrayrulewidth}{.5pt}
\setlength{\tabcolsep}{3pt}
\begin{table}[!t]
\centering
\footnotesize
\caption{Code complexity and efficiency assessment template.}
\label{tab:complexity_judging_template}
\vspace{-1em}
    \begin{tabular*}{\linewidth}{p{.99\linewidth}}
    \toprule
Evaluate the solutions and code provided by the assistant for their time efficiency and resource management. Assess how well the code optimizes computational time and resources while ensuring the accuracy and effectiveness of the implemented algorithms.

\textbf{Evaluation Criteria}: \\
\:\: \textit{Time Efficiency}: Does the code minimize computational time? \\
\:\: \textit{Resource Efficiency}: Does the code use resources (e.g., memory, CPU) judiciously? \\
\:\: \textit{Algorithm Effectiveness}: Are the chosen algorithms accurate and efficient in achieving the desired outcomes? \\
\:\: \textit{Optimization}: Has the code been optimized for quick processing without compromising the solution's correctness or efficiency? \\
\textbf{Scoring}: Rate outputs on a scale of 1 to 5: \\
\:\:1. \textit{Inefficient}: The code is resource-heavy and slow, with little evidence of optimization for time efficiency. \\
\:\:2. \textit{Somewhat Efficient}: The code shows some effort towards efficiency, but significant improvements are needed in time and resource management. \\
\:\:3. \textit{Moderately Efficient}: The code balances time and resource use reasonably well, with effective algorithm selection, but there's room for optimization. \\
\:\:4. \textit{Highly Efficient}: The code demonstrates strong time and resource efficiency, with well-chosen algorithms that provide swift and accurate results. \\
\:\:5. \textit{Optimally Efficient}: The code exemplifies the best practices in time and resource efficiency, with optimal algorithm selection and execution that ensure maximum effectiveness with minimal resource expenditure. \\

    \arrayrulecolor{black}
    \bottomrule
    \end{tabular*}
\end{table}

\renewcommand{\arraystretch}{1}
\setlength{\arrayrulewidth}{.5pt}
\setlength{\tabcolsep}{3pt}
\begin{table}[!t]
\centering
\footnotesize
\caption{Code readability assessment template.}
\label{tab:readability_judging_template_ap}
\vspace{-1em}
    \begin{tabular*}{\linewidth}{p{.99\linewidth}}
    \toprule
Evaluate the readability of code segments. Assess how comments and documentation contribute to understanding the code's logic, purpose, and operation.

\textbf{Evaluation Criteria}: \\
\:\: \textit{Clarity}: How clear and understandable are the code and its accompanying comments/documentation? \\
\:\: \textit{Conciseness}: Are the comments and documentation succinct yet informative? \\
\:\: \textit{Relevance}: Do the comments and documentation directly contribute to explaining the code's logic, objectives, and functionality? \\
\:\: \textit{Comprehensibility}: Can users of varying technical backgrounds easily grasp the code's purpose and how it works? \\
\textbf{Scoring}: Rate outputs on a scale of 1 to 5: \\
\:\:1. \textit{Poor Readability}: The code is hard to follow, with little to no helpful comments/documentation. \\
\:\:2. \textit{Basic Readability}: The code has minimal comments/documentation, offering limited clarity or insight. \\
\:\:3. \textit{Good Readability}: The code is reasonably clear with comments/documentation that aid understanding, though some areas could be improved. \\
\:\:4. \textit{Very Good Readability}: The code and comments/documentation are clear and concise, making the code's logic and purpose easily understandable. \\
\:\:5. \textit{Excellent Readability}: The code exemplifies outstanding readability, with clear, concise, and comprehensive comments/documentation that make it accessible to all users. \\

    \arrayrulecolor{black}
    \bottomrule
    \end{tabular*}
\end{table}

\renewcommand{\arraystretch}{1}
\setlength{\arrayrulewidth}{.5pt}
\setlength{\tabcolsep}{3pt}
\begin{table}[!t]
\centering
\footnotesize
\caption{Coding style assessment template.}
\label{tab:style_judging_template}
\vspace{-1em}
    \begin{tabular*}{\linewidth}{p{.99\linewidth}}
    \toprule
Evaluate the coding style of provided code segments. Assess how well the code adheres to the best practices of the language, focusing on readability, maintainability, and efficiency in line with the language's idiomatic style.

\textbf{Evaluation Criteria}: \\
\:\: \textit{Readability}: Is the code easy to read and understand? \\
\:\: \textit{Maintainability}: Can the code be easily modified or extended? \\
\:\: \textit{Efficiency}: Does the code execute tasks in an efficient manner? \\
\:\: \textit{Adherence to Idiomatic Style}: Does the code follow the stylistic norms and conventions specific to the programming language? \\
\textbf{Scoring}: Rate outputs on a scale of 1 to 5: \\
\:\:1. \textit{Non-Adherent}: The code largely ignores language conventions, resulting in poor readability and maintainability. \\
\:\:2. \textit{Somewhat Adherent}: The code makes some effort to follow language conventions but lacks consistency in readability, maintainability, or efficiency. \\
\:\:3. \textit{Moderately Adherent}: The code is generally in line with language conventions, offering fair readability and maintainability, with room for improvement in efficiency. \\
\:\:4. \textit{Highly Adherent}: The code strongly adheres to language conventions, demonstrating good readability, maintainability, and efficiency. \\
\:\:5. \textit{Exemplary Adherent}: The code exemplifies the best practices of the language, excelling in readability, maintainability, efficiency, and idiomatic style. \\

    \arrayrulecolor{black}
    \bottomrule
    \end{tabular*}
\end{table}

\renewcommand{\arraystretch}{1}
\setlength{\arrayrulewidth}{.5pt}
\setlength{\tabcolsep}{3pt}
\begin{table}[!t]
\centering
\footnotesize
\caption{System prompt for single-answer grading.}
\label{tab:system_prompt}
\vspace{-1em}
    \begin{tabular*}{\linewidth}{p{.99\linewidth}}
    \toprule
Please act as an impartial judge and evaluate the quality of the response provided by an AI assistant to the user instruction displayed below.
You will be given a Reference Response and the Assistant's Response. Begin your evaluation by comparing the Assistant's Response with the Reference Response. 
Your overall evaluation needs to be reflective of the specified Evaluation Criteria. Be as objective as possible. 
After providing your rationale, you must rate the Assistant's Response on a scale of 1 to 10.
DO NOT give a response to the instruction, ONLY provide your rationale followed by the rating. \\
    \arrayrulecolor{black}
    \bottomrule
    \end{tabular*}
\end{table}

\renewcommand{\arraystretch}{1}
\setlength{\arrayrulewidth}{.5pt}
\setlength{\tabcolsep}{3pt}
\begin{table}[!t]
\centering
\footnotesize
\caption{Instruction following template for single-answer grading.}
\label{tab:bench_instruction-following}
\vspace{-1em}
    \begin{tabular*}{\linewidth}{p{.99\linewidth}}
    \toprule
\# Instruction Following Assessment \\
\\
Evaluation Criteria: \\
\:\:- Precision in Following Instructions: Does the assistant adhere to the specifics of the provided instructions? \\
\:\:- Justification for Deviations: If deviations occur, are they justified by critical necessity or explicit user request? \\
\:\:- Alignment with User Directives: How well do the assistant's responses  match the user’s specified needs and expectations? \\
\:\:- Necessity of Deviations: Are any deviations from instructions made only in situations deemed absolutely necessary or upon direct user request? \\
\\
\#\# Format: \\
\\
\#\#\# Instruction \\
\:\:\texttt{[Instruction]} \\
\\
\#\#\# Reference Response: \\
\:\:\texttt{[Reference]} \\
\\
\#\#\# Assistant's Response: \\
\:\:\texttt{[Response]} \\
\\
\#\#\# Output \\
Rationale: \texttt{[Rationale for the rating in short sentences]} \\
Rating: \texttt{[Rating for Assistant's Response]} \\

--- \\

\#\# Example of output: \\
Rationale: The assistant's response ... \\
Rating: 2/10 \\

---

\#\# Annotation \\
\\
\#\#\# Instruction \\
\:\:\{instruction\} \\
\\
\#\#\# Reference Response: \\
\:\:\{reference\} \\
\\
\#\#\# Assistant's Response: \\
\:\:\{response\} \\
\\
\#\#\# Output \\

    \arrayrulecolor{black}
    \bottomrule
    \end{tabular*}
\end{table}

\renewcommand{\arraystretch}{1}
\setlength{\arrayrulewidth}{.5pt}
\setlength{\tabcolsep}{3pt}
\begin{table}[!t]
\centering
\footnotesize
\caption{Code explanation template for single-answer grading.}
\label{tab:bench_explanation}
\vspace{-1em}
    \begin{tabular*}{\linewidth}{p{.99\linewidth}}
    \toprule
\# Code Explanation Quality Assessment \\
\\
Evaluation Criteria: \\
\:\:- Clarity: How easy is it to understand the explanation? \\
\:\:- Depth: Does the explanation cover the logic, structure, and decisions behind the code? \\
\:\:- Relevance: Is the explanation relevant to the code's purpose and design philosophy? \\
\:\:- Accessibility: Can a broad audience understand the explanation, regardless of their technical background? \\
\\
\#\# Format: \\
\\
\#\#\# Instruction \\
\:\:\texttt{[Instruction]} \\
\\
\#\#\# Reference Response: \\
\:\:\texttt{[Reference]} \\
\\
\#\#\# Assistant's Response: \\
\:\:\texttt{[Response]} \\
\\
\#\#\# Output \\
Rationale: \texttt{[Rationale for the rating in short sentences]} \\
Rating: \texttt{[Rating for Assistant's Response]} \\

--- \\

\#\# Example of output: \\
Rationale: The assistant's response ... \\
Rating: 2/10 \\

---

\#\# Annotation \\
\\
\#\#\# Instruction \\
\:\:\{instruction\} \\
\\
\#\#\# Reference Response: \\
\:\:\{reference\} \\
\\
\#\#\# Assistant's Response: \\
\:\:\{response\} \\
\\
\#\#\# Output \\

    \arrayrulecolor{black}
    \bottomrule
    \end{tabular*}
\end{table}

\renewcommand{\arraystretch}{1}
\setlength{\arrayrulewidth}{.5pt}
\setlength{\tabcolsep}{3pt}
\begin{table}[!t]
\centering
\footnotesize
\caption{Code complexity and efficiency template for single-answer grading.}
\label{tab:bench_complexity}
\vspace{-1em}
    \begin{tabular*}{\linewidth}{p{.99\linewidth}}
    \toprule
\# Code Complexity and Efficiency Assessment \\
\\
Evaluation Criteria: \\
\:\:- Time Efficiency: Does the code minimize computational time? \\
\:\:- Resource Efficiency: Does the code use resources (e.g., memory, CPU) judiciously? \\
\:\:- Algorithm Effectiveness: Are the chosen algorithms accurate and efficient in achieving the desired outcomes? \\
\:\:- Optimization: Has the code been optimized for quick processing without compromising the solution's correctness or efficiency? \\
\\
\#\# Format: \\
\\
\#\#\# Instruction \\
\:\:\texttt{[Instruction]} \\
\\
\#\#\# Reference Response: \\
\:\:\texttt{[Reference]} \\
\\
\#\#\# Assistant's Response: \\
\:\:\texttt{[Response]} \\
\\
\#\#\# Output \\
Rationale: \texttt{[Rationale for the rating in short sentences]} \\
Rating: \texttt{[Rating for Assistant's Response]} \\

--- \\

\#\# Example of output: \\
Rationale: The assistant's response ... \\
Rating: 2/10 \\

---

\#\# Annotation \\
\\
\#\#\# Instruction \\
\:\:\{instruction\} \\
\\
\#\#\# Reference Response: \\
\:\:\{reference\} \\
\\
\#\#\# Assistant's Response: \\
\:\:\{response\} \\
\\
\#\#\# Output \\

    \arrayrulecolor{black}
    \bottomrule
    \end{tabular*}
\end{table}

\renewcommand{\arraystretch}{1}
\setlength{\arrayrulewidth}{.5pt}
\setlength{\tabcolsep}{3pt}
\begin{table}[!t]
\centering
\footnotesize
\caption{Code readability template for single-answer grading.}
\label{tab:bench_readability}
\vspace{-1em}
    \begin{tabular*}{\linewidth}{p{.99\linewidth}}
    \toprule
\# Code Readability Assessment \\
\\
Evaluation Criteria: \\
\:\:- Clarity: How clear and understandable are the code and its accompanying comments/documentation? \\
\:\:- Conciseness: Are the comments and documentation succinct yet informative? \\
\:\:- Relevance: Do the comments and documentation directly contribute to explaining the code's logic, objectives, and functionality? \\
\:\:- Comprehensibility: Can users of varying technical backgrounds easily grasp the code's purpose and how it works? \\
\\
\#\# Format: \\
\\
\#\#\# Instruction \\
\:\:\texttt{[Instruction]} \\
\\
\#\#\# Reference Response: \\
\:\:\texttt{[Reference]} \\
\\
\#\#\# Assistant's Response: \\
\:\:\texttt{[Response]} \\
\\
\#\#\# Output \\
Rationale: \texttt{[Rationale for the rating in short sentences]} \\
Rating: \texttt{[Rating for Assistant's Response]} \\

--- \\

\#\# Example of output: \\
Rationale: The assistant's response ... \\
Rating: 2/10 \\

---

\#\# Annotation \\
\\
\#\#\# Instruction \\
\:\:\{instruction\} \\
\\
\#\#\# Reference Response: \\
\:\:\{reference\} \\
\\
\#\#\# Assistant's Response: \\
\:\:\{response\} \\
\\
\#\#\# Output \\

    \arrayrulecolor{black}
    \bottomrule
    \end{tabular*}
\end{table}

\renewcommand{\arraystretch}{1}
\setlength{\arrayrulewidth}{.5pt}
\setlength{\tabcolsep}{3pt}
\begin{table}[!t]
\centering
\footnotesize
\caption{Coding style template for single-answer grading.}
\label{tab:bench_style}
\vspace{-1em}
    \begin{tabular*}{\linewidth}{p{.99\linewidth}}
    \toprule
\# Coding Style Assessment \\
\\
Evaluation Criteria: \\
\:\:- Readability: Is the code easy to read and understand? \\
\:\:- Maintainability: Can the code be easily modified or extended? \\
\:\:- Efficiency: Does the code execute tasks in an efficient manner? \\
\:\:- Adherence to Idiomatic Style: Does the code follow the stylistic norms and conventions specific to the programming language? \\
\\
\#\# Format: \\
\\
\#\#\# Instruction \\
\:\:\texttt{[Instruction]} \\
\\
\#\#\# Reference Response: \\
\:\:\texttt{[Reference]} \\
\\
\#\#\# Assistant's Response: \\
\:\:\texttt{[Response]} \\
\\
\#\#\# Output \\
Rationale: \texttt{[Rationale for the rating in short sentences]} \\
Rating: \texttt{[Rating for Assistant's Response]} \\

--- \\

\#\# Example of output: \\
Rationale: The assistant's response ... \\
Rating: 2/10 \\

---

\#\# Annotation \\
\\
\#\#\# Instruction \\
\:\:\{instruction\} \\
\\
\#\#\# Reference Response: \\
\:\:\{reference\} \\
\\
\#\#\# Assistant's Response: \\
\:\:\{response\} \\
\\
\#\#\# Output \\

    \arrayrulecolor{black}
    \bottomrule
    \end{tabular*}
\end{table}

\renewcommand{\arraystretch}{1}
\setlength{\arrayrulewidth}{.5pt}
\setlength{\tabcolsep}{3pt}
\begin{table}[!t]
\centering
\footnotesize
\caption{Evaluation template for HumanEval samples.}
\label{tab:humaneval_rating}
\vspace{-1em}
    \begin{tabular*}{\linewidth}{p{.99\linewidth}}
    \toprule
\# Functional Correctness Assessment \\
\\
Evaluation Criteria: \\
\:\:- Functional Correctness: Does the code correctly solve the problem as described in the instruction, passing all test cases? \\
\:\:- Practical Usefulness: Is the response executable and aligned with the intended task requirements? \\
\:\:- Clarity: Does the code offer a clear and concise solution without unnecessary complexity? \\
\\
\#\# Format: \\
\\
\#\#\# Instruction \\
\:\:\texttt{[Instruction]} \\
\\
\#\#\# Tests: \\
\:\:\texttt{[List of unit tests]} \\
\\
\#\#\# Assistant's Response: \\
\:\:\texttt{[Response]} \\
\\
\#\#\# Output \\
Rationale: \texttt{[Rationale for the rating in short sentences]} \\
Rating: \texttt{[Rating for Assistant's Response]} \\

--- \\

\#\# Examples of output: \\
Rationale: The code passes all test cases and solves the problem effectively with a concise implementation. \\
Rating: 10/10

Rationale: The code fails on key test cases and includes unnecessary complexity, making it difficult to understand. \\
Rating: 3/10

---

\#\# Annotation \\
\\
\#\#\# Instruction \\
\:\:\{instruction\} \\
\\
\#\#\# Reference Response: \\
\:\:\{reference\} \\
\\
\#\#\# Assistant's Response: \\
\:\:\{response\} \\
\\
\#\#\# Output \\

    \arrayrulecolor{black}
    \bottomrule
    \end{tabular*}
\end{table}

\renewcommand{\arraystretch}{1}
\setlength{\arrayrulewidth}{.5pt}
\setlength{\tabcolsep}{3pt}
\begin{table}[!t]
\centering
\footnotesize
\caption{Instruction following principles.}
\label{tab:principles_instruction-following}
\vspace{-1em}
    \begin{tabular*}{\linewidth}{p{.99\linewidth}}
    \toprule
    \arrayrulecolor{black!50}
The assistant is required to strictly follow the provided instructions, making sure that all responses and actions are in exact accordance with user directives. Only in critical situations or upon direct user request should it deviate from these instructions. \\
\midrule
It's essential for the assistant to rigorously stick to the instructions given, aligning its responses and actions accurately with user demands. Deviations are permissible solely when indispensable or specifically asked for by the user. \\
\midrule
The assistant must ensure precise compliance with instructions, aligning its responses and actions directly with user requirements. Any deviation should occur only if absolutely necessary or if explicitly requested by the user. \\
\midrule
In following instructions, the assistant should adhere closely to what is asked, ensuring responses and actions match the user’s specified needs. Departure from these instructions is allowed only when critically needed or when the user explicitly allows it. \\
\midrule
The assistant is obliged to faithfully follow the instructions provided, ensuring its responses and actions mirror user expectations precisely. Only in necessary circumstances or at the user's express request should it deviate. \\
\midrule
The assistant needs to maintain strict adherence to user instructions, aligning its responses and actions specifically with the user’s requests. It should deviate from these instructions only when there is an absolute necessity or upon user’s explicit instruction. \\
\midrule
It is crucial for the assistant to be diligent in following user instructions, keeping its responses and actions in tight alignment with what is requested. Deviations should be considered only under essential conditions or if requested by the user. \\
\midrule
The assistant's responses and actions must conform closely to the instructions given by the user. It should only diverge from these instructions in situations of utmost necessity or when explicitly asked by the user. \\
\midrule
The assistant should execute tasks in strict accordance with user instructions, ensuring all actions and responses are directly in line with user requirements. Only in pressing situations or with clear user consent should it deviate. \\
\midrule
For the assistant, precise adherence to the given instructions is mandatory, with responses and actions reflecting user requests accurately. Deviating from these guidelines is acceptable only in critical instances or when the user specifically requests it. \\
    \arrayrulecolor{black}
    \bottomrule
    \end{tabular*}
\end{table}

\renewcommand{\arraystretch}{1}
\setlength{\arrayrulewidth}{.5pt}
\setlength{\tabcolsep}{3pt}
\begin{table}[!t]
\centering
\footnotesize
\caption{Code explanation principles.}
\label{tab:principles_explanation}
\vspace{-1em}
    \begin{tabular*}{\linewidth}{p{.99\linewidth}}
    \toprule
    \arrayrulecolor{black!50}
The assistant is expected to provide detailed and accessible explanations for its code, clarifying the reasoning, architecture, and decisions involved, thereby aiding users in comprehending both the function and the rationale behind the code. \\
\midrule
It's important for the assistant to deliver comprehensive and intelligible explanations for the code, breaking down the logic, framework, and choices, to help users grasp the code's purpose and construction. \\
\midrule
The assistant must ensure that its code comes with clear, in-depth explanations, illuminating the underlying logic, structure, and choices, enabling users to not only understand the code's workings but also its design philosophy. \\
\midrule
In providing code, the assistant should also offer exhaustive and lucid explanations, elucidating the thought process, organization, and decisions behind the code, thus facilitating user understanding and application. \\
\midrule
The assistant should accompany its code with explicit and detailed explanations, shedding light on the logic, configuration, and decisions, helping users to comprehend and utilize the code effectively. \\
\midrule
For every piece of code, the assistant must provide clear, comprehensive explanations, detailing the logic, structure, and reasoning, enabling users to fully understand and apply the code in their contexts. \\
\midrule
The assistant is obliged to offer clear-cut and extensive explanations with its code, dissecting the logic, structure, and strategic choices, ensuring users gain a complete understanding of the code's purpose and design. \\
\midrule
Alongside its code, the assistant should present thorough and straightforward explanations, clarifying the underlying logic, framework, and choices, aiding users in understanding and adapting the code efficiently. \\
\midrule
The assistant must ensure that each code segment is accompanied by transparent and thorough explanations, unraveling the logic, structure, and choices, thus empowering users to grasp and leverage the code for their needs. \\
\midrule
It is essential for the assistant to couple its code with lucid, detailed explanations, unfolding the logic, composition, and decisions within the code, to help users not only understand its functionality but also the rationale behind its creation. \\

    \arrayrulecolor{black}
    \bottomrule
    \end{tabular*}
\end{table}

\renewcommand{\arraystretch}{1}
\setlength{\arrayrulewidth}{.5pt}
\setlength{\tabcolsep}{3pt}
\begin{table}[!t]
\centering
\footnotesize
\caption{Code complexity and efficiency principles.}
\label{tab:principles_complexity}
\vspace{-1em}
    \begin{tabular*}{\linewidth}{p{.99\linewidth}}
    \toprule
    \arrayrulecolor{black!50}
The assistant is tasked with delivering solutions and code that prioritize time efficiency, focusing on reducing computational duration and resource expenditure while ensuring the accuracy and effectiveness of the algorithms used. \\
\midrule
It's essential for the assistant to offer solutions and code that are optimized for quick processing, designed to minimize the time and resources required for execution, while maintaining correctness and efficiency. \\
\midrule
The assistant must ensure that its solutions and code are developed with optimal time efficiency in mind, aiming to lessen computational time and resource use, and selecting algorithms that are both accurate and swift in execution. \\
\midrule
In providing solutions, the assistant should focus on achieving the highest time efficiency, minimizing the amount of computational time and resources needed, while ensuring the selected algorithms are both correct and efficient. \\
\midrule
The assistant's goal should be to present solutions and code that excel in time efficiency, designed to curtail the duration of computation and resource usage, while adhering to algorithms that are effective and precise. \\
\midrule
For every solution and piece of code, the assistant is expected to prioritize time efficiency, striving to reduce computational time and resources, and favoring methods that are not only accurate but also swift. \\
\midrule
The assistant needs to concentrate on offering solutions and code that are time-efficient, minimizing computational and resource demands, while choosing algorithms that are both effective and efficient. \\
\midrule
In its solutions and code, the assistant should aim for optimal time efficiency, designing methods that cut down on computational time and resource use, while ensuring the approaches used are both correct and swift. \\
\midrule
The assistant is required to prioritize time efficiency in its solutions and code, focusing on minimizing the time and resources involved in computation, and selecting algorithms that are not only accurate but also quick. \\
\midrule
The assistant's objective should be to provide solutions and code that excel in time efficiency, aiming to reduce both computational time and resource consumption, while utilizing algorithms that are effective and efficient in their operation. \\

    \arrayrulecolor{black}
    \bottomrule
    \end{tabular*}
\end{table}

\renewcommand{\arraystretch}{1}
\setlength{\arrayrulewidth}{.5pt}
\setlength{\tabcolsep}{3pt}
\begin{table}[!t]
\centering
\footnotesize
\caption{Code readability principles.}
\label{tab:principles_readability}
\vspace{-1em}
    \begin{tabular*}{\linewidth}{p{.99\linewidth}}
    \toprule
    \arrayrulecolor{black!50}
The assistant must ensure its code and responses are marked by ease of understanding, incorporating straightforward, brief comments and documentation that clarify the code's logic, purpose, and operational process, facilitating user comprehension and learning. \\
\midrule
It is crucial for the assistant to make its code and responses readily comprehensible, using clear and concise explanations in comments and documentation to illuminate the reasoning, objectives, and functions of the code, thus aiding in user education and application. \\
\midrule
The assistant should provide code and responses that are easily graspable, with lucid, succinct comments and documentation elucidating the underlying logic, goals, and functionality of the code, enhancing the user's ability to follow and absorb the material. \\
\midrule
In delivering code and responses, the assistant is expected to prioritize understandability, featuring clear, compact comments and documentation that detail the code’s logic, intent, and capabilities, thereby assisting users in their learning journey. \\
\midrule
The assistant is tasked with ensuring that both its code and responses are straightforward to comprehend, integrating clear, brief comments and documentation that explicate the code's reasoning, purpose, and mechanisms, thus streamlining the learning process for users. \\
\midrule
For effective readability, the assistant should ensure its code and responses are effortlessly intelligible, employing clear, concise comments and documentation that convey the logic, objectives, and functionality of the code, simplifying the learning curve for users. \\
\midrule
The assistant's code and responses need to be accessible in terms of understanding, necessitating the use of clear, brief comments and documentation that outline the logic, intent, and function of the code, thereby enhancing user comprehension and learning. \\
\midrule
The assistant has the responsibility to ensure its code and responses are straightforward to understand, necessitating lucid, concise comments and documentation that outline the code’s reasoning, purpose, and operational aspects, key to ensuring users can easily follow and benefit from the provided material. \\
\midrule
The assistant should maintain a high level of readability in both its code and responses, achieved by integrating clear, straightforward comments and documentation that illuminate the code’s reasoning, purpose, and functionality, to assist users in easily grasping and learning from the content. \\
\midrule
In providing code and responses, the assistant must focus on making them easily understandable, entailing the inclusion of lucid, succinct comments and documentation that explicate the code’s logic, aim, and functional aspects, essential for ensuring users can easily follow and benefit from the provided content. \\

    \arrayrulecolor{black}
    \bottomrule
    \end{tabular*}
\end{table}

\renewcommand{\arraystretch}{1}
\setlength{\arrayrulewidth}{.5pt}
\setlength{\tabcolsep}{3pt}
\begin{table}[!t]
\centering
\footnotesize
\caption{Coding style principles.}
\label{tab:principles_style}
\vspace{-1em}
    \begin{tabular*}{\linewidth}{p{.99\linewidth}}
    \toprule
    \arrayrulecolor{black!50}
The assistant should write code that exemplifies the best practices of the language used, focusing on readability, maintainability, and efficiency in line with the language's idiomatic style. \\
\midrule
It is important for the assistant to adopt a coding style that is characteristic of the language, balancing clarity, conciseness, and efficiency to produce easily understandable and efficient code. \\
\midrule
The assistant's coding approach should reflect the standard conventions of the language, prioritizing a style that is both effective and in harmony with the language’s best practices. \\
\midrule
In coding, the assistant must aim for a style that is not only syntactically correct but also adheres to the idiomatic nuances of the programming language, ensuring efficiency and readability. \\
\midrule
The assistant is expected to craft code that follows the established stylistic norms of the programming language, optimizing for a balance between clarity, efficiency, and language-specific best practices. \\
\midrule
The assistant should maintain a coding style that aligns with the language's conventional practices, ensuring that the code is efficient, readable, and easily maintained. \\
\midrule
Coding in a style that is representative of the language's community standards, the assistant should strive for code that is succinct, efficient, and easily interpreted by others. \\
\midrule
The assistant's coding should demonstrate a clear adherence to the stylistic principles of the language, focusing on creating code that is both high-performing and elegantly written. \\
\midrule
In its coding practices, the assistant should embody the ideal balance of efficiency, readability, and adherence to the language's stylistic conventions, ensuring optimal code performance. \\
\midrule
The assistant's coding style should be a reflection of the language’s standard practices, emphasizing a clean, efficient, and idiomatic approach to ensure both performance and readability. \\

    \arrayrulecolor{black}
    \bottomrule
    \end{tabular*}
\end{table}

\end{document}